\documentclass[twocolumn,trackchanges,resetfootnote]{aastex701}
\usepackage{graphicx}
\usepackage{subcaption}
\usepackage{float} 
\usepackage{amsmath}
\usepackage{booktabs}
\usepackage{array}  % Para definir columnas centradas
\usepackage{amssymb} % Para \checkmark
\usepackage{longtable}
\usepackage{enumitem}

\shorttitle{Physical Properties of AGN-Hosting Galaxy Mergers}
\shortauthors{Troncoso et al.}

\graphicspath{{./}{figures/}}

\begin{document}

\title{BASS LIV. Physical Properties of AGN-Hosting Galaxy Mergers from Multiwavelength SED Fitting}

\author[0009-0002-2740-2993]{Marco Troncoso}
\affiliation{Instituto de Astrofísica, Facultad de Física, Pontificia Universidad Católica de Chile, Av. Vicuña Mackenna 4860, 782-0436 Macul, Santiago, Chile}
\email[show]{marco.troncoso@uc.cl}

\author[0000-0001-7568-6412]{Ezequiel Treister}
\affiliation{Instituto de Alta Investigaci{\'{o}}n, Universidad de Tarapac{\'{a}}, Casilla 7D, Arica, Chile}
\email{etreister@academicos.uta.cl}

\author[0000-0003-0006-8681]{Alejandra Rojas}
\affiliation{Departamento de F\'isica, Universidad T\'ecnica Federico Santa Mar\'ia, Vicu\~{n}a Mackenna 3939, San Joaqu\'in, Santiago de Chile, Chile}
\email{ale.rojaslilayu@gmail.com}

\author[0000-0003-0946-6176]{Médéric Boquien}
\affiliation{Université Côte d’Azur, Observatoire de la Côte d’Azur, CNRS, Laboratoire Lagrange, 06000, Nice, France}
\email{mederic.boquien@oca.eu}

\author[0000-0002-8686-8737]{Franz Bauer}
\affiliation{Instituto de Alta Investigaci{\'{o}}n, Universidad de Tarapac{\'{a}}, Casilla 7D, Arica, Chile}
\email{franz.e.bauer@gmail.com}

\author[0000-0002-7998-9581]{Michael J. Koss}
\affiliation{Eureka Scientific, 2452 Delmer Street, Suite 100, Oakland, CA 94602-3017, USA}
\email{mike.koss@eurekasci.com} 

\author[0000-0002-9508-3667]{Roberto J. Assef}
\affiliation{Instituto de Estudios Astrof\'isicos, Facultad de Ingenier\'ia y Ciencias, Universidad Diego Portales, Av. Ej\'ercito Libertador 441, Santiago, Chile}
\email{roberto.assef@mail.udp.cl} 

\author[0000-0001-5649-7798]{Miguel Parra Tello}
\affiliation{Instituto de Astrofísica, Facultad de Física, Pontificia Universidad Católica de Chile, Av. Vicuña Mackenna 4860, 782-0436 Macul, Santiago, Chile}
\email{mgparra@uc.cl}

\author[0000-0001-8931-1152]{Ignacio del Moral-Castro}
\affiliation{Instituto de Astrofísica, Facultad de Física, Pontificia Universidad Católica de Chile, Av. Vicuña Mackenna 4860, 782-0436 Macul, Santiago, Chile}
\email{ignaciodelmoralcastro.astro@gmail.com}

\author[0000-0001-5231-2645]{Claudio Ricci}
\affiliation{Instituto de Estudios Astrof\'isicos, Facultad de Ingenier\'ia y Ciencias, Universidad Diego Portales, Av. Ej\'ercito Libertador 441, Santiago, Chile}
\email{claudio.ricci.astro@gmail.com}

\author[0000-0002-7928-416X]{Sophia Dai}
\affiliation{National Astronomical Observatories, Chinese Academy of Sciences, Beijing 100101, China}
\email{daysophia@gmail.com}

\author[0000-0002-5037-951X]{Kyuseok Oh}
\affiliation{Korea Astronomy and Space Science Institute, Daedeokdae-ro 776, Yuseong-gu, Daejeon 34055, Republic of Korea}
\email{oh@kasi.re.kr}

\author[0000-0001-5742-5980]{Federica Ricci}
\affiliation{Roma Tre University, Rome 3169070, Italy}
\email{fed.ricci89@gmail.com}

\author[0000-0001-5742-5980]{Alessandro Peca}
\affiliation{Eureka Scientific, 2452 Delmer Street, Suite 100, Oakland, CA 94602-3017, USA}
\email{peca.alessandro@gmail.com}

\author[0000-0002-0745-9792]{C. Megan Urry}
\affiliation{Yale Center for Astronomy \& Astrophysics and Department of Physics, Yale University, P.O. Box 208120, New Haven, CT
06520-8120, USA}
\email{meg.urry@yale.edu}

\author[0009-0007-9018-1077]{Kriti Kamal Gupta}
\affiliation{STAR Institute, Li\`ege Universit\'e, Quartier Agora - All\'ee du six Ao\^ut, 19c B-4000 Li\`ege, Belgium}
\affiliation{Sterrenkundig Observatorium, Universiteit Gent, Krijgslaan 281 S9, B-9000 Gent, Belgium}
\email{kriti.gupta@mail.udp.cl}

\author[0000-0001-8349-3055]{Giacomo Venturi}
\affiliation{Scuola Normale Superiore, Piazza dei Cavalieri 7, I-56126 Pisa, Italy}
\affiliation{INAF - Osservatorio Astrofisico di Arcetri, Largo E. Fermi 5, I-50125 Firenze, Italy}
\email{giacomo.venturi1@sns.it}

\author[0000-0002-8177-6905]{Matilde Signorini}
\affiliation{European Space Research and Technology Centre, Noordwijk-Binnen 2749812, The Netherlands}
\email{matilde.signorini@uniroma3.it}

\author[0000-0002-7962-5446]{Richard Mushotzky}
\affiliation{Department of Astronomy, University of Maryland, College Park, MD 20742, USA}
\email{rmushotz@umd.edu}

\author[0000-0002-1233-9998]{David Sanders}
\affiliation{Institute for Astronomy, 2680 Woodlawn Drive, University of Hawaii, Honolulu, HI 96822, USA}
\email{sanders@ifa.hawaii.edu}

\begin{abstract}

Galaxy mergers are believed to play an important role in triggering rapid supermassive black hole (SMBH) growth. As merging nuclei approach each other, the physical properties of the participating galaxies and the associated SMBH growth are expected to evolve significantly. This study measures and characterizes these physical properties throughout the merger sequence. We constructed multiwavelength Spectral Energy Distributions (SEDs) from hard X-rays to the far-infrared (FIR) for a sample of 72 nearby Active Galactic Nuclei (AGN) host galaxies. The sample comprises 64 interacting systems, including single AGNs in mergers and dual AGNs, with nuclear separations $\leq$30 kpc, as well as eight isolated active galaxies with merging features. We carefully adapted available photometric measurements at each wavelength to account for their complex morphologies and varying spatial resolutions, to perform SED fitting using CIGALE, aimed to derive critical physical properties. Our results reveal that merging galaxies hosting AGN(s) show deviations from the star-forming main sequence, and a wide range of star formation rates (SFRs). Both AGN activity and star formation are significantly influenced by the merger process, but these effects are more prominent in major, mass ratios $<$4:1, interactions. We find that the projected nuclear separation is not a good tracer of the merger stage. Instead, morphological classification accurately assesses the merger progression. Based on this morphological analysis, late-stage mergers exhibit elevated SFRs (5.1$\times$), AGN luminosities (2.4$\times$), and nuclear obscuration (2.8$\times$) compared to earlier stages, supporting previous findings and reinforcing the link between merger-driven galaxy evolution and SMBH growth.

\end{abstract}


\section{Introduction} \label{sec:intro}

Understanding the formation and growth of supermassive black holes (SMBHs) across cosmic time remains among the most critical challenges in extragalactic astronomy. Observational evidence shows that the properties of galaxies strongly correlate with those of the SMBHs in their nuclei \citep{1995ARA&A..33..581K, 1998AJ....115.2285M, 2000ApJ...539L...9F, 2000ApJ...539L..13G, 2009ApJ...698..198G, 2009ApJ...704.1135B}, suggesting a fundamental link between galaxy evolution and SMBH growth. Active Galactic Nuclei (AGN) are found in galaxies where material is being accreted onto the central black hole at high rates \citep{1969Natur.223..690L}. However, for this material to reach the vicinity of the black hole, it must lose about 99\% of its angular momentum \citep[][and references therein]{2011MNRAS.415.1027H}. Simulations indicate that large-scale galaxy mergers of two or more galaxies provide an efficient mechanism for the gas to cease equilibrium and fall towards the SMBH, triggering the AGN \citep{2005Natur.433..604D, 2006ApJS..163....1H, 2008ApJS..175..356H}. Observationally, the AGN-merger connection is well-established \citep{2019MNRAS.487.2491E, 2024A&A...690A.326L}, and in particular these mergers seem to be responsible for the most luminous AGNs \citep{2012ApJ...758L..39T} and Quasars \citep{1988ApJ...325...74S, 2016ApJ...822L..32F, 2018Sci...362.1034D}, implying that such collisions substantially accelerate the SMBH growth, although contradictory results were reported, where luminous AGNs do not originate from major galaxy mergers \citep[e.g.,][]{2016ApJ...830..156M} or even are not associated with merging systems \citep[e.g.,][]{2019MNRAS.489..497Z}. Studying and finding these objects is crucial because the processes occurring before the final coalescence of SMBH remain largely unknown, and could potentially represent a significant fraction, up to $\sim$60\%, of the total SMBH growth \citep{2012ApJ...758L..39T, 2015MNRAS.451.2517S}. 

In the final stages of the merging process, when both nuclei of the merging galaxies approach each other (at separations of approximately $10\ \mathrm{kpc}$, although the exact division is rather arbitrary), these systems are more likely to host an AGN \citep[e.g.,][]{2010ApJ...716L.125K, 2011ApJ...739...57K, 2011MNRAS.418.2043E,2021ApJ...923...36S}. This finding suggests that, at those close separations, the gas and dust are efficiently transported toward the galactic center, thereby fueling the SMBH growth. This is also seen in simulations \citep[e.g.,][]{1991ApJ...370L..65B, 2024MNRAS.528.5864B}. 

Observations have shown that merger-triggered AGNs are likely to become heavily obscured by gas and dust \citep[e.g.,][]{2017MNRAS.468.1273R, 2019ApJ...875..117P, 2021MNRAS.506.5935R}, particularly at later stages of the process. Additionally, galaxy merger simulations further predict that along with this obscuration, a peak in SMBH accretion occurs when the two galactic nuclei are near each other prior to coalescence \citep[e.g.,][]{2018MNRAS.478.3056B}. Due to the strong effect of obscuration, many AGNs in merging galaxies remain undetected at optical to even soft X-ray ($<$10 keV) wavelengths. However, a sizable fraction can be found through hard X-ray observations, which are capable of penetrating dense gas and dust columns, and are therefore less affected by obscuration \citep{2015ApJ...815L..13R, 2022ApJS..261....9A}. Notable surveys employing this approach include Swift/BAT \citep{2018ApJS..235....4O} and NuSTAR \citep{2016ApJ...831..185H}. 

%Among available tools, hard X-ray observations have proven especially effective, as they can penetrate thick gas and dust columns and are less affected from star formation, which also contributes to the IR emission.

%\citep{2014ApJ...789..112C}

The majority of AGNs are found in isolated, non-merging galaxies and are referred to as single AGNs. However, AGNs can also reside in merging systems. When only one nucleus hosts an AGN, the system is often referred to as a single AGN in a merger (single AGN iM), while when both nuclei are active, the system is defined as a dual AGN \citep[e.g.,][]{2009ApJ...702L..82C, 2012ApJ...746L..22K}. Observations have shown that the fraction of dual AGNs increases as the merger progresses \citep[e.g.,][]{2012ApJ...746L..22K, 2018ApJ...856...93F}. However, it remains uncertain whether these phenomena represent different stages of the merger process \citep{2023MNRAS.524.4482B} or both nuclei eventually become active at other times due to the disparity in time scales between mergers and AGN activity ($t_\mathrm{AGN} \ll t_\mathrm{merger}$; \citealp{2012ApJ...748L...7V}), along with an intrinsic AGN variability that could change the accretion rate of the source from minutes to decades \citep[e.g.,][]{2007MNRAS.375..989W}. 

The triggering of AGN activity is not the only consequence of galaxy mergers. The accumulation of matter in the inner regions of galaxies during these events can also lead to elevated star formation rates \citep[SFRs; e.g.,][]{2005ApJ...620L..79S}. It is well established that the formation of massive galaxies begins with an intense phase of star formation in the early epochs \citep[$SFR_{MS}$ at $z<2$  e.g.,][]{2017ApJ...839...26D}. Over time, the available gas is gradually consumed, and the galaxy settles into a more stable star-forming phase, with SFRs typical of massive, star-forming galaxies defined as the MS, which are generally classified as \textit{late-type galaxies} \citep[e.g.,][]{2011A&A...533A.119E}. However, when these galaxies undergo a merger, the interaction can trigger a new starburst event in the nuclear region of the system \citep[$>$ 4 $\times\ SFR_{MS}$;  e.g.,][]{1988ApJ...325...74S, 2011ApJ...739L..40R, 2015A&A...575A..74S, 2023ApJS..265...37Y, 2024MNRAS.527.2037R}, significantly enhancing the SFR in both progenitor galaxies \citep[e.g.,][]{2021ApJ...923....6J}. Eventually, the merger-driven starburst exhausts the cold gas reservoir and quenches star formation, producing a massive, passive remnant. These systems, known as \textit{early-type galaxies}, are now understood to be the final outcome of the merger process and its associated evolutionary phases \citep[e.g.,][]{2002ApJS..143..315V, 2005ApJ...620L..79S, 2014MNRAS.440..889S}.

Many previous studies have investigated merging galaxies using imaging techniques \citep[e.g.,][]{1988ApJ...325...74S, 2008AJ....135.1877E, 2011MNRAS.418.2043E, 2012MNRAS.426..549S, 2021ApJ...923....6J} and simulations \citep[e.g.,][]{1996ApJ...464..641M, 2008A&A...492...31D, 2008ApJS..175..356H, 2024MNRAS.528.5864B}. However, these studies have not focused on the sources with the closest separations, particularly in those hosting an active nucleus (or two). Detecting \textit{mid-} and \textit{late-stage mergers} hosting AGNs requires high spatial resolution data to resolve closely separated nuclei, making angular resolution crucial for such studies. Additionally, as previously noted, nuclear obscuration plays a key role in identifying these systems. Therefore, observations at wavelengths less affected by dust obscuration, such as the MIR or X-ray bands, are also essential. A summary of all AGNs hosted by galaxies involved in mergers and how they were identified is presented in \cite{2024arXiv241112799P}.

Measuring the physical properties of such sources is challenging due to the vast spectral range typically covered by the AGN emission, spanning from X-rays originating in the corona, UV/optical from the accretion disk, IR from the dusty torus, to mm emission also partly from the corona. This wide wavelength coverage creates a highly degenerate problem to characterize the AGN emission and separate it from the host galaxy, particularly when data coverage and depth are limited. To address this, assembling multi-band photometric data and fitting spectral energy distributions (SEDs) offers a robust approach to disentangling the different emission components of a galaxy and accurately determining its physical properties. This method has been employed in various studies for AGN sources \citep[e.g.,][]{1986ApJ...308...59E, 2007ApJ...663...81P, 2022ApJ...927..192Y}.

Merging galaxies are expected to exhibit distinct physical properties compared to non-merging systems \citep[e.g.,][]{2008AJ....135.1877E}. These properties are believed to evolve throughout the merger process, with particularly strong changes observed at projected separations below $30\ \rm kpc$, such as AGN activity \citep[e.g.,][]{2011MNRAS.418.2043E}, AGN obscuration \citep[e.g.,][]{2014MNRAS.441.1297S}, and elevated SFRs \citep[e.g.,][]{2013MNRAS.433L..59P}. Therefore, studying merging systems at these separations provides crucial insights into how the interaction during merger stages shapes the nuclear activity, and the resulting physical characteristics of the galaxies involved \citep{2012MNRAS.426..549S}.

In this paper, we investigate how galaxy mergers impact the physical properties of SMBHs and their host galaxies across different stages of the merger process. To achieve this, we analyze a sample of merging systems with projected nuclear separations spanning up to $30\ \rm kpc$, using multi-band SED fitting with wavelength coverage from hard X-rays to the FIR. Section \ref{sec:sample} describes the sample selection process. In Section \ref{sec: photometry}, we present the wavelength bands used and explain how photometric fluxes are measured for SED fitting. The detailed procedure for fitting SEDs using CIGALE \citep{2019A&A...622A.103B} is outlined in Section \ref{sec: sed}. The derived physical parameters are presented and analyzed in Section \ref{sec: results}, providing insight into the effects of mergers on SMBHs and their host galaxies. Finally, a discussion of the results and our conclusions are presented in Sections \ref{sec:discussion} and \ref{sec: conclusion}, respectively. Throughout this paper, we assume a flat $\Lambda$CDM cosmology with $H_0 = 70\ \mathrm{km\,s^{-1}\,Mpc^{-1}}$, $\Omega_\mathrm{m} = 0.3$, and $\Omega_\Lambda = 0.7$.

%Code Investigating GALaxy Emission (

%During the interaction, the participating galaxies are subject to dynamic friction, reducing their relative motion until they reach a separation of approximately 1 kpc. At this point, the galaxies may begin to merge their stellar bulges.

%To conserve angular momentum, stars in the vicinity of the SMBHs are ejected in a process known as \textit{Stellar Hardening}. On smaller scales, a similar process affects the surrounding gas. As the SMBHs draw closer, they begin to emit gravitational waves to conserve energy, ultimately bringing them into closer proximity (as illustrated in Figure 5 of \cite{2023ApJ...942L..24K}). 

%What happens next is still uncertain. It is believed that the SMBHs spend an extended period in close orbit before finally merging, a stage known as the \textit{Final Parsec Problem}. Although some simulations have explored this phenomenon \citep{2009MNRAS.398.1392L, 2024PhLB..85638908K}, the precise mechanisms remain unclear. Identifying systems with the closest nuclei provides valuable opportunity to compare observations with these theoretical models.

%\vspace{-1cm}

\section{Sample} \label{sec:sample}

%\clearpage
 
The sources analyzed in this work were selected from the Swift/BAT spectroscopic survey\footnote{\url{https://www.bass-survey.com}} (BASS, \citealp{2022ApJS..261....1K}), which comprises the most luminous hard X-ray selected AGNs in the local universe. From this parent sample, we selected a subset of merging systems with small radial velocity differences \citep[$<300 km/s$;][]{2012ApJ...746L..22K} in the nearby universe ($z$$<$0.1), where the relatively small distances allow us to spatially resolve the nuclei of the participating galaxies to small physical separations. 

\begin{figure}[htpb]
    \centering
    \includegraphics[width=\columnwidth]{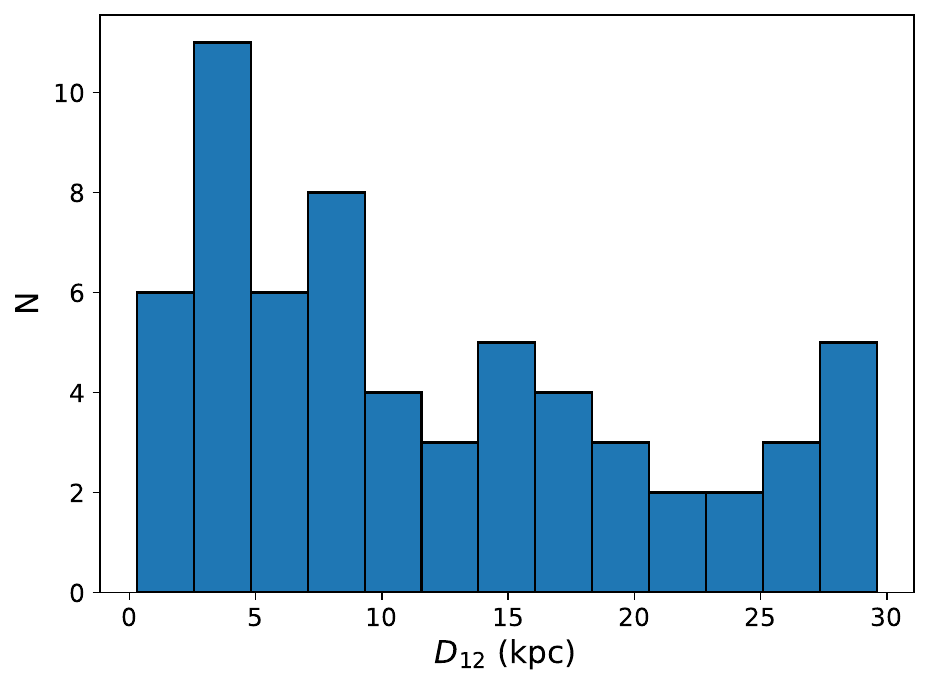}
    \caption{Distribution of projected nuclear separation in kpc, for the 64 galaxies involved in merging systems in our sample; 35/64 galaxies are late-stage mergers ($D_{12} < 10\ \mathrm{kpc}$).} 
    \label{fig:sep_dist}
\end{figure}

To identify these systems, we use three complementary approaches. First, we selected candidates for late-stage mergers that might have projected nuclear separations of less than 10 kpc. These sources were required to exhibit merging features (e.g., disturbed disks and/or tidal tails) and/or have a close companion. To better resolve these systems, they were observed with the Gemini South Adaptive Optics Imager (GSAOI; \citealp{2004SPIE.5492.1033M}), a near-IR adaptive optics instrument installed on the Gemini Multi-Conjugate Adaptive Optics System (GeMS) at Gemini South in northern Chile. GSAOI has a pixel size of $0.02''$/pix and consists of four 2048 x 2048 chips with $2''$ gaps, providing a field of view of $85'' \times 85''$. Using GSAOI, 22 sources were observed in the H ($1.64\ \rm \mu m$) and Ks ($2.16\ \rm \mu m$) bands between December 2021 and December 2022 (GS-2021B-Q-132, GS-2022A-Q-140, GS-2022B-Q-234). These images were reduced with THELI v3 \citep{2013ApJS..209...21S}. This dataset enabled us to measure the projected separation of the merging systems or assess whether the sources were involved in mergers. We identified four sources with separations below $10\ \mathrm{kpc}$, 10 mergers with larger separations, and eight sources not involved in mergers but exhibiting merger features.

The second selection comes from the sample of BASS AGNs studied by \cite{2018Natur.563..214K}, who reported the identification of 17 late-stage mergers. This subsample includes mergers with separations ranging from $0.3\ \mathrm{kpc}$ to $10\ \mathrm{kpc}$.

The third selection is based on a sample of mergers with already measured separations between $4\ \mathrm{kpc}$ and $28\ \mathrm{kpc}$, thereby expanding the range of separations studied in this work. These galaxies were selected from the merger sample identified by \cite{2012ApJ...746L..22K} using the 70-month BAT catalog \citep{2013ApJS..207...19B}, later expanded to include the southern hemisphere and the 105-month sample \citep{2018ApJS..235....4O}. These objects were also observed in the millimeter with the Atacama Compact Array (ACA; \citealp{2009PASJ...61....1I}) under project ID 2023.1.01471.S. This subsample contributes 33 sources to the final sample.

The parameters used to measure the projected nuclear separations in kpc are the redshift and projected separation in arcsec. (1) The redshifts used in this work were primarily measured from the [OIII] emission line in the NLR \citep{2022ApJS..261....2K}. For the BASS sample, this line has, in general, a very high signal-to-noise ratio. The spectra were obtained with resolutions of R $\approx$ 1,200–6,000 using Palomar/DBSP and VLT/X-shooter, corresponding to velocity uncertainties of only 20-50 km/s. In redshift space, at z $\approx$ 0.03, this translates to an associated uncertainty of $\sim$0.3$\%$. Such small uncertainties have a negligible impact on the projected separations. (2) Projected separations (in arcsec) are determined from the centroids of the galaxies. The main source of uncertainty is therefore how accurately the centroids can be measured, which is limited by the angular resolution of the imaging and improves with increasing S/N, which for the nuclear regions is usually high. For the GSAOI sample, where the separations were measured directly in this work, the angular resolution is $\sim$0.1$"$, yielding well-resolved centroids with negligible positional uncertainties. For the \cite{2018Natur.563..214K} late-stage merger sample ($\mathrm{D_{12}}<10\ \mathrm{kpc}$), which relies on Keck AO imaging (resolution $\sim$0.17$"$), the situation is similar. For the larger merger sample, in the worst-case images from DES, SDSS, or LEGACY (resolution $\sim$1.0$"$) were used for this measurement, but even in these cases, the projected separations are mostly above 10$"$, so the relative uncertainty remains negligible.

In total, our data set includes 72 sources, which are summarized in Table \ref{tab:agn_merger}. 64 are merging systems spanning projected separations ranging from $0.3$ to $29.6\ \mathrm{kpc}$. Figure \ref{fig:sep_dist} shows the projected separation distribution in the merging sample, defined as the distance between the two nuclei projected onto the plane of the sky, measured in $\mathrm{kpc}$. Among them, nine sources are confirmed dual AGNs based on mm and/or X-ray observations \citep{2012ApJ...746L..22K}, and 55 are classified as single AGNs in mergers, where only one AGN is identified. 

\begin{figure*}[htbp]
    \centering
    \includegraphics[width=\textwidth]{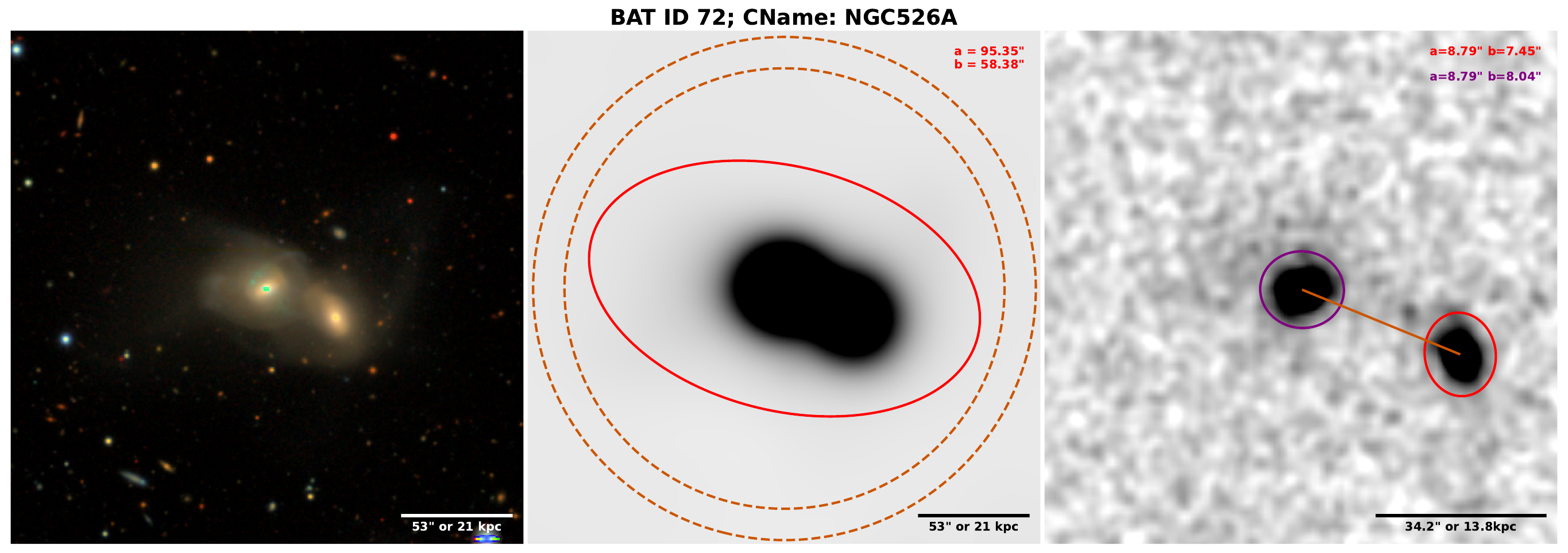}
    \caption{Left: RGB image for BAT ID: 72 constructed from the g,r, and z filters. Middle: Example convolved image presenting the aperture and annulus used for the photometry. Right: NIR image from 2MASS in the Ks band, showing the apertures for the M1/M2 measurement and the projected separation indicated by the scale bar. The complete figure set (71 images) is available in the online journal.} 
    \label{fig: source}
\end{figure*}

We estimated stellar mass ratios (M$_1$/M$_2$) using NIR fluxes as a proxy. In particular, we use the K filter for these measurements: 2MASS for separated sources, and GSAOI and Keck for closer companions. In some particular cases, we used the HST F160W filter and the VHS K band. All images used are presented in Figure \ref{fig: source}.

For the systems in our sample, we adopted a 4:1 mass ratio as the dividing boundary between major and minor mergers \citep[e.g.,][]{1972ApJ...178..623T, 2005AJ....129..682H, 2006ApJ...638..686C}. Based on this criterion, 33 systems can be classified as major mergers and 31 as minor mergers. Additionally, eight galaxies from the GSAOI sample exhibit merger features without a visible companion, suggesting they may be in an advanced merger stage, may have already coalesced, or may not be involved in mergers. Figure \ref{fig: source} present an example, and the full subset of sources is illustrated in %\input{fig2set}
\figsetstart
\figsetnum{2}
\figsettitle{Source Images.}

\figsetgrpstart
\figsetgrpnum{2.1}
\figsetgrptitle{13}
\figsetplot{sources/BAT13.pdf}
\figsetgrpnote{Left: RGB image for BAT ID: 13 constructed using g,r, and z filters. Middle: Convolved image with aperture and annulus used for photometry. Right: NIR image from Keck II in the Kp band, showing the apertures for the M1/M2 measurement and the projected separation indicated by the scale bar.}
\figsetgrpend

\figsetgrpstart
\figsetgrpnum{2.2}
\figsetgrptitle{28}
\figsetplot{sources/BAT28.pdf}
\figsetgrpnote{Left: RGB image for BAT ID: 28 constructed using g,r, and z filters. Middle: Convolved image with aperture and annulus used for photometry. Right: NIR image from 2MASS in the Ks band, showing the apertures for the M1/M2 measurement and the projected separation indicated by the scale bar.}
\figsetgrpend

\figsetgrpstart
\figsetgrpnum{2.3}
\figsetgrptitle{44}
\figsetplot{sources/BAT44.pdf}
\figsetgrpnote{Left: RGB image for BAT ID: 44 constructed using g,r, and z filters. Middle: Convolved image with aperture and annulus used for photometry. Right: NIR image from 2MASS in the Ks band, showing the apertures for the M1/M2 measurement and the projected separation indicated by the scale bar.}
\figsetgrpend

\figsetgrpstart
\figsetgrpnum{2.4}
\figsetgrptitle{60}
\figsetplot{sources/BAT60.pdf}
\figsetgrpnote{Left: RGB image for BAT ID: 60 constructed using g,r, and z filters. Middle: Convolved image with aperture and annulus used for photometry. Right: NIR image from Keck II in the Kp band, showing the apertures for the M1/M2 measurement and the projected separation indicated by the scale bar.}
\figsetgrpend

\figsetgrpstart
\figsetgrpnum{2.5}
\figsetgrptitle{63}
\figsetplot{sources/BAT63.pdf}
\figsetgrpnote{Left: RGB image for BAT ID: 63 constructed using g,r, and z filters. Middle: Convolved image with aperture and annulus used for photometry. Right: NIR image from 2MASS in the Ks band, showing the apertures for the M1/M2 measurement and the projected separation indicated by the scale bar.}
\figsetgrpend

\figsetgrpstart
\figsetgrpnum{2.6}
\figsetgrptitle{72}
\figsetplot{sources/BAT72.pdf}
\figsetgrpnote{Left: RGB image for BAT ID: 72 constructed using g,r, and z filters. Middle: Convolved image with aperture and annulus used for photometry. Right: NIR image from 2MASS in the Ks band, showing the apertures for the M1/M2 measurement and the projected separation indicated by the scale bar.}
\figsetgrpend

\figsetgrpstart
\figsetgrpnum{2.7}
\figsetgrptitle{73}
\figsetplot{sources/BAT73.pdf}
\figsetgrpnote{Left: RGB image for BAT ID: 73 constructed using g,r, and z filters. Middle: Convolved image with aperture and annulus used for photometry. Right: NIR image from 2MASS in the Ks band, showing the apertures for the M1/M2 measurement and the projected separation indicated by the scale bar.}
\figsetgrpend

\figsetgrpstart
\figsetgrpnum{2.8}
\figsetgrptitle{83}
\figsetplot{sources/BAT83.pdf}
\figsetgrpnote{Left: RGB image for BAT ID: 83 constructed using g,r, and z filters. Middle: Convolved image with aperture and annulus used for photometry. Right: NIR image from 2MASS in the Ks band, showing the apertures for the M1/M2 measurement and the projected separation indicated by the scale bar.}
\figsetgrpend

\figsetgrpstart
\figsetgrpnum{2.9}
\figsetgrptitle{88}
\figsetplot{sources/BAT88.pdf}
\figsetgrpnote{Left: RGB image for BAT ID: 88 constructed using g,r, and z filters. Middle: Convolved image with aperture and annulus used for photometry. Right: NIR image from Keck II in the Kp band, showing the apertures for the M1/M2 measurement and the projected separation indicated by the scale bar.}
\figsetgrpend

\figsetgrpstart
\figsetgrpnum{2.10}
\figsetgrptitle{89}
\figsetplot{sources/BAT89.pdf}
\figsetgrpnote{Left: RGB image for BAT ID: 89 constructed using g,r, and z filters. Middle: Convolved image with aperture and annulus used for photometry. Right: NIR image from GSAOI in the Kp band, showing the apertures for the M1/M2 measurement and the projected separation indicated by the scale bar.}
\figsetgrpend

\figsetgrpstart
\figsetgrpnum{2.11}
\figsetgrptitle{134}
\figsetplot{sources/BAT134.pdf}
\figsetgrpnote{Left: RGB image for BAT ID: 134 constructed using g,r, and z filters. Middle: Convolved image with aperture and annulus used for photometry. Right: NIR image from HST in the F160W band, showing the apertures for the M1/M2 measurement and the projected separation indicated by the scale bar.}
\figsetgrpend

\figsetgrpstart
\figsetgrpnum{2.12}
\figsetgrptitle{136}
\figsetplot{sources/BAT136.pdf}
\figsetgrpnote{Left: RGB image for BAT ID: 136 constructed using g,r, and z filters. Middle: Convolved image with aperture and annulus used for photometry. Right: NIR image from 2MASS in the Ks band, showing the apertures for the M1/M2 measurement and the projected separation indicated by the scale bar.}
\figsetgrpend

\figsetgrpstart
\figsetgrpnum{2.13}
\figsetgrptitle{157}
\figsetplot{sources/BAT157.pdf}
\figsetgrpnote{Left: RGB image for BAT ID: 157 constructed using g,r, and z filters. Middle: Convolved image with aperture and annulus used for photometry. Right: NIR image from 2MASS in the Ks band, showing the apertures for the M1/M2 measurement and the projected separation indicated by the scale bar.}
\figsetgrpend

\figsetgrpstart
\figsetgrpnum{2.14}
\figsetgrptitle{159}
\figsetplot{sources/BAT159.pdf}
\figsetgrpnote{Left: RGB image for BAT ID: 159 constructed using g,r, and z filters. Middle: Convolved image with aperture and annulus used for photometry. Right: NIR image from VHS in the K band, showing the apertures for the M1/M2 measurement and the projected separation indicated by the scale bar.}
\figsetgrpend

\figsetgrpstart
\figsetgrpnum{2.15}
\figsetgrptitle{189}
\figsetplot{sources/BAT189.pdf}
\figsetgrpnote{Left: RGB image for BAT ID: 189 constructed using g,r, and z filters. Middle: Convolved image with aperture and annulus used for photometry. Right: NIR image from 2MASS in the Ks band, showing the apertures for the M1/M2 measurement and the projected separation indicated by the scale bar.}
\figsetgrpend

\figsetgrpstart
\figsetgrpnum{2.16}
\figsetgrptitle{193}
\figsetplot{sources/BAT193.pdf}
\figsetgrpnote{Left: RGB image for BAT ID: 193 constructed using g,r, and z filters. Middle: Convolved image with aperture and annulus used for photometry. Right: NIR image from 2MASS in the Ks band, showing the apertures for the M1/M2 measurement and the projected separation indicated by the scale bar.}
\figsetgrpend

\figsetgrpstart
\figsetgrpnum{2.17}
\figsetgrptitle{197}
\figsetplot{sources/BAT197.pdf}
\figsetgrpnote{Left: RGB image for BAT ID: 197 constructed using g,r, and z filters. Middle: Convolved image with aperture and annulus used for photometry. Right: NIR image from VHS in the K band, showing the apertures for the M1/M2 measurement and the projected separation indicated by the scale bar.}
\figsetgrpend

\figsetgrpstart
\figsetgrpnum{2.18}
\figsetgrptitle{217}
\figsetplot{sources/BAT217.pdf}
\figsetgrpnote{Left: RGB image for BAT ID: 217 constructed using g,r, and z filters. Middle: Convolved image with aperture and annulus used for photometry. Right: NIR image from 2MASS in the Ks band, showing the apertures for the M1/M2 measurement and the projected separation indicated by the scale bar.}
\figsetgrpend

\figsetgrpstart
\figsetgrpnum{2.19}
\figsetgrptitle{218}
\figsetplot{sources/BAT218.pdf}
\figsetgrpnote{Left: RGB image for BAT ID: 218 constructed using g,r, and z filters. Middle: Convolved image with aperture and annulus used for photometry. Right: NIR image from Keck II in the Kp band, showing the apertures for the M1/M2 measurement and the projected separation indicated by the scale bar.}
\figsetgrpend

\figsetgrpstart
\figsetgrpnum{2.20}
\figsetgrptitle{231}
\figsetplot{sources/BAT231.pdf}
\figsetgrpnote{Left: RGB image for BAT ID: 231 constructed using g,r, and z filters. Middle: Convolved image with aperture and annulus used for photometry. Right: NIR image from 2MASS in the Ks band, showing the apertures for the M1/M2 measurement and the projected separation indicated by the scale bar.}
\figsetgrpend

\figsetgrpstart
\figsetgrpnum{2.21}
\figsetgrptitle{243}
\figsetplot{sources/BAT243.pdf}
\figsetgrpnote{Left: RGB image for BAT ID: 243 constructed using g,r, and z filters. Middle: Convolved image with aperture and annulus used for photometry. Right: NIR image from GSAOI in the Kp band, showing the apertures for the M1/M2 measurement and the projected separation indicated by the scale bar.}
\figsetgrpend

\figsetgrpstart
\figsetgrpnum{2.22}
\figsetgrptitle{246}
\figsetplot{sources/BAT246.pdf}
\figsetgrpnote{Left: RGB image for BAT ID: 246 constructed using g,r, and z filters. Middle: Convolved image with aperture and annulus used for photometry. Right: NIR image from 2MASS in the Ks band, showing the apertures for the M1/M2 measurement and the projected separation indicated by the scale bar.}
\figsetgrpend

\figsetgrpstart
\figsetgrpnum{2.23}
\figsetgrptitle{260}
\figsetplot{sources/BAT260.pdf}
\figsetgrpnote{Left: RGB image for BAT ID: 260 constructed using g,r, and z filters. Middle: Convolved image with aperture and annulus used for photometry. Right: NIR image from 2MASS in the Ks band, showing the apertures for the M1/M2 measurement and the projected separation indicated by the scale bar.}
\figsetgrpend

\figsetgrpstart
\figsetgrpnum{2.24}
\figsetgrptitle{280}
\figsetplot{sources/BAT280.pdf}
\figsetgrpnote{Left: RGB image for BAT ID: 280 constructed using g,r, and z filters. Middle: Convolved image with aperture and annulus used for photometry. Right: NIR image from 2MASS in the Ks band, showing the apertures for the M1/M2 measurement and the projected separation indicated by the scale bar.}
\figsetgrpend

\figsetgrpstart
\figsetgrpnum{2.25}
\figsetgrptitle{302}
\figsetplot{sources/BAT302.pdf}
\figsetgrpnote{Left: RGB image for BAT ID: 302 constructed using g,r, and z filters. Middle: Convolved image with aperture and annulus used for photometry. Right: NIR image from 2MASS in the Ks band, showing the apertures for the M1/M2 measurement and the projected separation indicated by the scale bar.}
\figsetgrpend

\figsetgrpstart
\figsetgrpnum{2.26}
\figsetgrptitle{303}
\figsetplot{sources/BAT303.pdf}
\figsetgrpnote{Left: RGB image for BAT ID: 303 constructed using g,r, and z filters. Middle: Convolved image with aperture and annulus used for photometry. Right: NIR image from Keck II in the Kp band, showing the apertures for the M1/M2 measurement and the projected separation indicated by the scale bar.}
\figsetgrpend

\figsetgrpstart
\figsetgrpnum{2.27}
\figsetgrptitle{305}
\figsetplot{sources/BAT305.pdf}
\figsetgrpnote{Left: RGB image for BAT ID: 305 constructed using g,r, and z filters. Middle: Convolved image with aperture and annulus used for photometry. Right: NIR image from GSAOI in the Kp band, showing the apertures for the M1/M2 measurement and the projected separation indicated by the scale bar.}
\figsetgrpend

\figsetgrpstart
\figsetgrpnum{2.28}
\figsetgrptitle{316}
\figsetplot{sources/BAT316.pdf}
\figsetgrpnote{Left: RGB image for BAT ID: 316 constructed using g,r, and z filters. Middle: Convolved image with aperture and annulus used for photometry. Right: NIR image from 2MASS in the Ks band, showing the apertures for the M1/M2 measurement and the projected separation indicated by the scale bar.}
\figsetgrpend

\figsetgrpstart
\figsetgrpnum{2.29}
\figsetgrptitle{318}
\figsetplot{sources/BAT318.pdf}
\figsetgrpnote{Left: RGB image for BAT ID: 318 constructed using g,r, and z filters. Middle: Convolved image with aperture and annulus used for photometry. Right: NIR image from GSAOI in the Kp band, showing the apertures for the M1/M2 measurement and the projected separation indicated by the scale bar.}
\figsetgrpend

\figsetgrpstart
\figsetgrpnum{2.30}
\figsetgrptitle{329}
\figsetplot{sources/BAT329.pdf}
\figsetgrpnote{Left: RGB image for BAT ID: 329 constructed using g,r, and z filters. Middle: Convolved image with aperture and annulus used for photometry. Right: NIR image from 2MASS in the Ks band, showing the apertures for the M1/M2 measurement and the projected separation indicated by the scale bar.}
\figsetgrpend

\figsetgrpstart
\figsetgrpnum{2.31}
\figsetgrptitle{342}
\figsetplot{sources/BAT342.pdf}
\figsetgrpnote{Left: RGB image for BAT ID: 342 constructed using g,r, and z filters. Middle: Convolved image with aperture and annulus used for photometry. Right: NIR image from 2MASS in the Ks band, showing the apertures for the M1/M2 measurement and the projected separation indicated by the scale bar.}
\figsetgrpend

\figsetgrpstart
\figsetgrpnum{2.32}
\figsetgrptitle{405}
\figsetplot{sources/BAT405.pdf}
\figsetgrpnote{Left: RGB image for BAT ID: 405 constructed using g,r, and z filters. Middle: Convolved image with aperture and annulus used for photometry. Right: NIR image from Keck II in the Kp band, showing the apertures for the M1/M2 measurement and the projected separation indicated by the scale bar.}
\figsetgrpend

\figsetgrpstart
\figsetgrpnum{2.33}
\figsetgrptitle{416}
\figsetplot{sources/BAT416.pdf}
\figsetgrpnote{Left: RGB image for BAT ID: 416 constructed using g,r, and z filters. Middle: Convolved image with aperture and annulus used for photometry. Right: NIR image from VHS in the K band, showing the apertures for the M1/M2 measurement and the projected separation indicated by the scale bar.}
\figsetgrpend

\figsetgrpstart
\figsetgrpnum{2.34}
\figsetgrptitle{430}
\figsetplot{sources/BAT430.pdf}
\figsetgrpnote{Left: RGB image for BAT ID: 430 constructed using g,r, and z filters. Middle: Convolved image with aperture and annulus used for photometry. Right: NIR image from Keck II in the Kp band, showing the apertures for the M1/M2 measurement and the projected separation indicated by the scale bar.}
\figsetgrpend

\figsetgrpstart
\figsetgrpnum{2.35}
\figsetgrptitle{442}
\figsetplot{sources/BAT442.pdf}
\figsetgrpnote{Left: RGB image for BAT ID: 442 constructed using g,r, and z filters. Middle: Convolved image with aperture and annulus used for photometry. Right: NIR image from 2MASS in the Ks band, showing the apertures for the M1/M2 measurement and the projected separation indicated by the scale bar.}
\figsetgrpend

\figsetgrpstart
\figsetgrpnum{2.36}
\figsetgrptitle{465}
\figsetplot{sources/BAT465.pdf}
\figsetgrpnote{Left: RGB image for BAT ID: 465 constructed using g,r, and z filters. Middle: Convolved image with aperture and annulus used for photometry. Right: NIR image from 2MASS in the Ks band, showing the apertures for the M1/M2 measurement and the projected separation indicated by the scale bar.}
\figsetgrpend

\figsetgrpstart
\figsetgrpnum{2.37}
\figsetgrptitle{471}
\figsetplot{sources/BAT471.pdf}
\figsetgrpnote{Left: RGB image for BAT ID: 471 constructed using g,r, and z filters. Middle: Convolved image with aperture and annulus used for photometry. Right: NIR image from 2MASS in the Ks band, showing the apertures for the M1/M2 measurement and the projected separation indicated by the scale bar.}
\figsetgrpend

\figsetgrpstart
\figsetgrpnum{2.38}
\figsetgrptitle{489}
\figsetplot{sources/BAT489.pdf}
\figsetgrpnote{Left: RGB image for BAT ID: 489 constructed using g,r, and z filters. Middle: Convolved image with aperture and annulus used for photometry. Right: NIR image from 2MASS in the Ks band, showing the apertures for the M1/M2 measurement and the projected separation indicated by the scale bar.}
\figsetgrpend

\figsetgrpstart
\figsetgrpnum{2.39}
\figsetgrptitle{497}
\figsetplot{sources/BAT497.pdf}
\figsetgrpnote{Left: RGB image for BAT ID: 497 constructed using g,r, and z filters. Middle: Convolved image with aperture and annulus used for photometry. Right: NIR image from 2MASS in the Ks band, showing the apertures for the M1/M2 measurement and the projected separation indicated by the scale bar.}
\figsetgrpend

\figsetgrpstart
\figsetgrpnum{2.40}
\figsetgrptitle{533}
\figsetplot{sources/BAT533.pdf}
\figsetgrpnote{Left: RGB image for BAT ID: 533 constructed using g,r, and z filters. Middle: Convolved image with aperture and annulus used for photometry. Right: NIR image from Keck II in the Kp band, showing the apertures for the M1/M2 measurement and the projected separation indicated by the scale bar.}
\figsetgrpend

\figsetgrpstart
\figsetgrpnum{2.41}
\figsetgrptitle{543}
\figsetplot{sources/BAT543.pdf}
\figsetgrpnote{Left: RGB image for BAT ID: 543 constructed using g,r, and z filters. Middle: Convolved image with aperture and annulus used for photometry. Right: NIR image from Keck II in the Kp band, showing the apertures for the M1/M2 measurement and the projected separation indicated by the scale bar.}
\figsetgrpend

\figsetgrpstart
\figsetgrpnum{2.42}
\figsetgrptitle{552}
\figsetplot{sources/BAT552.pdf}
\figsetgrpnote{Left: RGB image for BAT ID: 552 constructed using g,r, and z filters. Middle: Convolved image with aperture and annulus used for photometry. Right: NIR image from Keck II in the Kp band, showing the apertures for the M1/M2 measurement and the projected separation indicated by the scale bar.}
\figsetgrpend

\figsetgrpstart
\figsetgrpnum{2.43}
\figsetgrptitle{557}
\figsetplot{sources/BAT557.pdf}
\figsetgrpnote{Left: RGB image for BAT ID: 557 constructed using g,r, and z filters. Middle: Convolved image with aperture and annulus used for photometry. Right: NIR image from Keck II in the Kp band, showing the apertures for the M1/M2 measurement and the projected separation indicated by the scale bar.}
\figsetgrpend

\figsetgrpstart
\figsetgrpnum{2.44}
\figsetgrptitle{567}
\figsetplot{sources/BAT567.pdf}
\figsetgrpnote{Left: RGB image for BAT ID: 567 constructed using g,r, and z filters. Middle: Convolved image with aperture and annulus used for photometry. Right: NIR image from 2MASS in the Ks band, showing the apertures for the M1/M2 measurement and the projected separation indicated by the scale bar.}
\figsetgrpend

\figsetgrpstart
\figsetgrpnum{2.45}
\figsetgrptitle{584}
\figsetplot{sources/BAT584.pdf}
\figsetgrpnote{Left: RGB image for BAT ID: 584 constructed using g,r, and z filters. Middle: Convolved image with aperture and annulus used for photometry. Right: NIR image from 2MASS in the Ks band, showing the apertures for the M1/M2 measurement and the projected separation indicated by the scale bar.}
\figsetgrpend

\figsetgrpstart
\figsetgrpnum{2.46}
\figsetgrptitle{605}
\figsetplot{sources/BAT605.pdf}
\figsetgrpnote{Left: RGB image for BAT ID: 605 constructed using g,r, and z filters. Middle: Convolved image with aperture and annulus used for photometry. Right: NIR image from Keck II in the Kp band, showing the apertures for the M1/M2 measurement and the projected separation indicated by the scale bar.}
\figsetgrpend

\figsetgrpstart
\figsetgrpnum{2.47}
\figsetgrptitle{606}
\figsetplot{sources/BAT606.pdf}
\figsetgrpnote{Left: RGB image for BAT ID: 606 constructed using g,r, and z filters. Middle: Convolved image with aperture and annulus used for photometry. Right: NIR image from 2MASS in the Ks band, showing the apertures for the M1/M2 measurement and the projected separation indicated by the scale bar.}
\figsetgrpend

\figsetgrpstart
\figsetgrpnum{2.48}
\figsetgrptitle{641}
\figsetplot{sources/BAT641.pdf}
\figsetgrpnote{Left: RGB image for BAT ID: 641 constructed using g,r, and z filters. Middle: Convolved image with aperture and annulus used for photometry. Right: NIR image from 2MASS in the Ks band, showing the apertures for the M1/M2 measurement and the projected separation indicated by the scale bar.}
\figsetgrpend

\figsetgrpstart
\figsetgrpnum{2.49}
\figsetgrptitle{669}
\figsetplot{sources/BAT669.pdf}
\figsetgrpnote{Left: RGB image for BAT ID: 669 constructed using g,r, and z filters. Middle: Convolved image with aperture and annulus used for photometry. Right: NIR image from 2MASS in the Ks band, showing the apertures for the M1/M2 measurement and the projected separation indicated by the scale bar.}
\figsetgrpend

\figsetgrpstart
\figsetgrpnum{2.50}
\figsetgrptitle{678}
\figsetplot{sources/BAT678.pdf}
\figsetgrpnote{Left: RGB image for BAT ID: 678 constructed using g,r, and z filters. Middle: Convolved image with aperture and annulus used for photometry. Right: NIR image from 2MASS in the Ks band, showing the apertures for the M1/M2 measurement and the projected separation indicated by the scale bar.}
\figsetgrpend

\figsetgrpstart
\figsetgrpnum{2.51}
\figsetgrptitle{703}
\figsetplot{sources/BAT703.pdf}
\figsetgrpnote{Left: RGB image for BAT ID: 703 constructed using g,r, and z filters. Middle: Convolved image with aperture and annulus used for photometry. Right: NIR image from HST in the F160W band, showing the apertures for the M1/M2 measurement and the projected separation indicated by the scale bar.}
\figsetgrpend

\figsetgrpstart
\figsetgrpnum{2.52}
\figsetgrptitle{712}
\figsetplot{sources/BAT712.pdf}
\figsetgrpnote{Left: RGB image for BAT ID: 712 constructed using g,r, and z filters. Middle: Convolved image with aperture and annulus used for photometry. Right: NIR image from 2MASS in the Ks band, showing the apertures for the M1/M2 measurement and the projected separation indicated by the scale bar.}
\figsetgrpend

\figsetgrpstart
\figsetgrpnum{2.53}
\figsetgrptitle{817}
\figsetplot{sources/BAT817.pdf}
\figsetgrpnote{Left: RGB image for BAT ID: 817 constructed using g,r, and z filters. Middle: Convolved image with aperture and annulus used for photometry. Right: NIR image from Keck II in the Kp band, showing the apertures for the M1/M2 measurement and the projected separation indicated by the scale bar.}
\figsetgrpend

\figsetgrpstart
\figsetgrpnum{2.54}
\figsetgrptitle{841}
\figsetplot{sources/BAT841.pdf}
\figsetgrpnote{Left: RGB image for BAT ID: 841 constructed using g,r, and z filters. Middle: Convolved image with aperture and annulus used for photometry. Right: NIR image from HST in the F160W band, showing the apertures for the M1/M2 measurement and the projected separation indicated by the scale bar.}
\figsetgrpend

\figsetgrpstart
\figsetgrpnum{2.55}
\figsetgrptitle{862}
\figsetplot{sources/BAT862.pdf}
\figsetgrpnote{Left: RGB image for BAT ID: 862 constructed using g,r, and z filters. Middle: Convolved image with aperture and annulus used for photometry. Right: NIR image from Keck II in the Kp band, showing the apertures for the M1/M2 measurement and the projected separation indicated by the scale bar.}
\figsetgrpend

\figsetgrpstart
\figsetgrpnum{2.56}
\figsetgrptitle{998}
\figsetplot{sources/BAT998.pdf}
\figsetgrpnote{Left: RGB image for BAT ID: 998 constructed using g,r, and z filters. Middle: Convolved image with aperture and annulus used for photometry. Right: NIR image from Keck II in the Kp band, showing the apertures for the M1/M2 measurement and the projected separation indicated by the scale bar.}
\figsetgrpend

\figsetgrpstart
\figsetgrpnum{2.57}
\figsetgrptitle{1063}
\figsetplot{sources/BAT1063.pdf}
\figsetgrpnote{Left: NIR image of BAT ID: 1063. Middle: Convolved image with aperture and annulus used for photometry. Right: NIR image from 2MASS in the Ks band, showing the apertures for the M1/M2 measurement and the projected separation indicated by the scale bar.}
\figsetgrpend

\figsetgrpstart
\figsetgrpnum{2.58}
\figsetgrptitle{1077}
\figsetplot{sources/BAT1077.pdf}
\figsetgrpnote{Left: RGB image for both sources from BAT ID: 1077 constructed using g,r, and z filters. Middle: Convolved image with aperture and annulus used for photometry. Right: NIR image from 2MASS in the Ks band, showing the apertures for the M1/M2 measurement and the projected separation indicated by the scale bar.}
\figsetgrpend

\figsetgrpstart
\figsetgrpnum{2.59}
\figsetgrptitle{10777}
\figsetplot{sources/BAT10777.pdf}
\figsetgrpnote{Left: RGB image for BAT ID: 1077 counterpart constructed using g,r, and z filters. Middle: Convolved image with aperture and annulus used for photometry. Right: NIR image from 2MASS in the Ks band, showing the apertures for the M1/M2 measurement and the projected separation indicated by the scale bar.}
\figsetgrpend

\figsetgrpstart
\figsetgrpnum{2.60}
\figsetgrptitle{1130}
\figsetplot{sources/BAT1130.pdf}
\figsetgrpnote{Left: RGB image for BAT ID: 1130 constructed using g,r, and z filters. Middle: Convolved image with aperture and annulus used for photometry. Right: NIR image from 2MASS in the Ks band, showing the apertures for the M1/M2 measurement and the projected separation indicated by the scale bar.}
\figsetgrpend

\figsetgrpstart
\figsetgrpnum{2.61}
\figsetgrptitle{1139}
\figsetplot{sources/BAT1139.pdf}
\figsetgrpnote{Left: RGB image for BAT ID: 1139 constructed using g,r, and z filters. Middle: Convolved image with aperture and annulus used for photometry. Right: NIR image from Keck II in the Kp band, showing the apertures for the M1/M2 measurement and the projected separation indicated by the scale bar.}
\figsetgrpend

\figsetgrpstart
\figsetgrpnum{2.62}
\figsetgrptitle{1141}
\figsetplot{sources/BAT1141.pdf}
\figsetgrpnote{Left: RGB image for BAT ID: 1141 constructed using g,r, and z filters. Middle: Convolved image with aperture and annulus used for photometry. Right: NIR image from 2MASS in the Ks band, showing the apertures for the M1/M2 measurement and the projected separation indicated by the scale bar.}
\figsetgrpend

\figsetgrpstart
\figsetgrpnum{2.63}
\figsetgrptitle{1182}
\figsetplot{sources/BAT1182.pdf}
\figsetgrpnote{Left: RGB image for BAT ID: 1182 constructed using g,r, and z filters. Middle: Convolved image with aperture and annulus used for photometry. Right: NIR image from 2MASS in the Ks band, showing the apertures for the M1/M2 measurement and the projected separation indicated by the scale bar.}
\figsetgrpend

\figsetgrpstart
\figsetgrpnum{2.64}
\figsetgrptitle{1210}
\figsetplot{sources/BAT1210.pdf}
\figsetgrpnote{Left: RGB image for BAT ID: 1210 constructed using g,r, and z filters. Middle: Convolved image with aperture and annulus used for photometry. Right: NIR image from 2MASS in the Ks band, showing the apertures for the M1/M2 measurement and the projected separation indicated by the scale bar.}
\figsetgrpend

\figsetgrpstart
\figsetgrpnum{2.65}
\figsetgrptitle{1255}
\figsetplot{sources/BAT1255.pdf}
\figsetgrpnote{Left: RGB image for BAT ID: 1255 constructed using g,r, and z filters. Middle: Convolved image with aperture and annulus used for photometry. Right: NIR image from 2MASS in the Ks band, showing the apertures for the M1/M2 measurement and the projected separation indicated by the scale bar.}
\figsetgrpend

\figsetgrpstart
\figsetgrpnum{2.66}
\figsetgrptitle{1262}
\figsetplot{sources/BAT1262.pdf}
\figsetgrpnote{Left: RGB image for BAT ID: 1262 constructed using g,r, and z filters. Middle: Convolved image with aperture and annulus used for photometry. Right: NIR image from 2MASS in the Ks band, showing the apertures for the M1/M2 measurement and the projected separation indicated by the scale bar.}
\figsetgrpend

\figsetgrpstart
\figsetgrpnum{2.67}
\figsetgrptitle{1356}
\figsetplot{sources/BAT1356.pdf}
\figsetgrpnote{Left: RGB image for BAT ID: 1356 constructed using g,r, and z filters. Middle: Convolved image with aperture and annulus used for photometry. Right: NIR image from 2MASS in the Ks band, showing the apertures for the M1/M2 measurement and the projected separation indicated by the scale bar.}
\figsetgrpend

\figsetgrpstart
\figsetgrpnum{2.68}
\figsetgrptitle{1390}
\figsetplot{sources/BAT1390.pdf}
\figsetgrpnote{Left: RGB image for BAT ID: 1390 constructed using g,r, and z filters. Middle: Convolved image with aperture and annulus used for photometry. Right: NIR image from 2MASS in the Ks band, showing the apertures for the M1/M2 measurement and the projected separation indicated by the scale bar.}
\figsetgrpend

\figsetgrpstart
\figsetgrpnum{2.69}
\figsetgrptitle{1426}
\figsetplot{sources/BAT1426.pdf}
\figsetgrpnote{Left: NIR image for BAT ID: 1426 constructed using g,r, and z filters. Middle: Convolved image with aperture and annulus used for photometry. Right: NIR image from 2MASS in the Ks band, showing the apertures for the M1/M2 measurement and the projected separation indicated by the scale bar.}
\figsetgrpend

\figsetgrpstart
\figsetgrpnum{2.70}
\figsetgrptitle{1439}
\figsetplot{sources/BAT1439.pdf}
\figsetgrpnote{Left: RGB image for BAT ID: 1439 constructed using g,r, and z filters. Middle: Convolved image with aperture and annulus used for photometry. Right: NIR image from GSAOI in the Kp band, showing the apertures for the M1/M2 measurement and the projected separation indicated by the scale bar.}
\figsetgrpend

\figsetgrpstart
\figsetgrpnum{2.71}
\figsetgrptitle{1627}
\figsetplot{sources/BAT1627.pdf}
\figsetgrpnote{Left: RGB image for BAT ID: 1627 constructed using g,r, and z filters. Middle: Convolved image with aperture and annulus used for photometry. Right: NIR image from 2MASS in the Ks band, showing the apertures for the M1/M2 measurement and the projected separation indicated by the scale bar.}
\figsetgrpend

\figsetend

\startlongtable
\begin{deluxetable*}{llllllllll}
\tablewidth{\textwidth}
\tablecaption{Sample of AGN-hosting galaxy mergers used in this work. \label{tab:agn_merger}}
\tablehead{
\colhead{ID} & \colhead{SWIFT} & \colhead{Object Name} & \colhead{$z$} & \colhead{$D_{12}$ ($\prime\prime$)} & \colhead{$D_{12}$ (kpc)} & \colhead{M1/M2} & \colhead{AGN Type} & \colhead{Merger AGN} & \colhead{Stage} \\
\colhead{(1)} & \colhead{(2)} & \colhead{(3)} & \colhead{(4)} & \colhead{(5)} & \colhead{(6)} & \colhead{(7)} & \colhead{(8)} & \colhead{(9)} & \colhead{(10)}
}
\startdata
\multicolumn{10}{c}{\textbf{Major Galaxy Mergers}}
\\ \hline
405                           & J0804.6+1045                     & UGC4211                                & 0.035                          & 0.3                                                  & 0.3                                       & 3.6                              & Sy2                                 & dual AGN                              & D                                \\
841                           & J1652.9+0223                     & NGC6240                                & 0.025                          & 1.8                                                  & 0.9                                       & 2.1                              & Sy2                                 & dual AGN                              & D                                \\
13                            & J0025.8+6818                     & LEDA136991                             & 0.012                          & 4.3                                                  & 1.1                                       & 1.3                              & Sy2                                 & single AGN iM                         & C                                \\
998                           & J1845.4+7211                     & Z341-6                                 & 0.047                          & 1.3                                                  & 1.2                                       & 1.7                              & Sy2                                 & dual AGN                              & D                                \\
88                            & J0138.8+2925                     & LEDA138434                             & 0.072                          & 1.2                                                  & 1.8                                       & 2.8                              & Sy1.9                               & single AGN iM                         & D                                \\
60                            & J0113.8+1313                     & Mrk975                                 & 0.05                           & 2.5                                                  & 2.5                                       & 2.3                              & Sy1                                 & single AGN iM                         & C                                \\
243                           & J0450.7-5813                     & RBS594                                 & 0.09                           & 1.5                                                  & 2.6                                       & 3.9                              & Sy1                                 & single AGN iM                         & D                                \\
817                           & J1631.7+2353                     & SDSSJ163115.52+235257.4                & 0.059                          & 2.3                                                  & 2.7                                       & 2.1                              & Sy1.9                               & dual AGN                              & C                                \\
552                           & J1136.0+2132                     & Mrk739E                                & 0.029                          & 6.1                                                  & 3.7                                       & 1.3                              & Sy1                                 & dual AGN                              & C                                \\
533                           & J1114.3+2020                     & 2MASXJ11140245+2023140                 & 0.027                          & 7.6                                                  & 4.1                                       & 2.8                              & Sy2                                 & single AGN iM                         & D                                \\
641                           & J1252.3-1323                     & NGC4748                                & 0.014                          & 15.7                                                 & 4.7                                       & 3.9                              & Sy1                                 & single AGN iM                         & C                                \\
157                           & J0255.2-0011                     & NGC1142                                & 0.029                          & 8.5                                                  & 4.8                                       & 2.5                              & Sy2                                 & single AGN iM                         & C                                \\
703                           & J1355.9+1822                     & Mrk463                                 & 0.05                           & 4.8                                                  & 4.9                                       & 3.6                              & Sy1.9                               & dual AGN                              & C                                \\
1139                          & J2207.3+1013                     & UGC11910                               & 0.027                          & 14.3                                                 & 7.9                                       & 2.3                              & Sy2                                 & single AGN iM                         & C                                \\
678                           & J1334.8-2328                     & ESO509-IG066                           & 0.034                          & 12.1                                                 & 8.1                                       & 2.5                              & Sy1.9                               & single AGN iM                         & B                                \\
497                           & J1023.5+1952                     & NGC3227                                & 0.003                          & 153.0                                                & 8.9                                       & 1.8                              & Sy1                                 & dual AGN                              & B                                \\
605                           & J1214.3+2933                     & Was49b                                 & 0.063                          & 7.1                                                  & 9.2                                       & 1.9                              & Sy1                                 & single AGN iM                         & C                                \\
862                           & J1708.6+2155                     & SDSSJ170859.13+215308.1                & 0.073                          & 6.7                                                  & 10.0                                      & 3.3                              & Sy1                                 & dual AGN                              & B                                \\
260                           & J0508.1+1727                     & 2MASXJ05081967+1721483                 & 0.017                          & 29.4                                                 & 10.6                                      & 2.8                              & Sy1.9                               & single AGN iM                         & A                                \\
72                            & J0123.8-3504                     & NGC526A                                & 0.019                          & 35.9                                                 & 13.8                                      & 2.0                              & Sy1.9                               & dual AGN                              & B                                \\
1356                          & J0941.2-3846                     & 2MASXJ09410102-3847509                 & 0.064                          & 11.9                                                 & 14.9                                      & 2.7                              & Sy2                                 & single AGN iM                         & A                                \\
567                           & J1145.6-1819                     & HE1143-1810                            & 0.033                          & 22.2                                                 & 14.9                                      & 2.9                              & Sy1                                 & single AGN iM                         & B                                \\
1141                          & J2209.1-2747                     & NGC7214                                & 0.023                          & 33.9                                                 & 15.6                                      & 4.0                              & Sy1                                 & single AGN iM                         & C                                \\
1262                          & J0324.9-0304                     & NGC1320                                & 0.008                          & 95.1                                                 & 15.6                                      & 1.7                              & Sy2                                 & single AGN iM                         & A                                \\
44                            & J0100.9-4750                     & 2MASXJ01003490-4752033                 & 0.048                          & 18.7                                                 & 17.7                                      & 2.7                              & Sy2                                 & single AGN iM                         & B                                \\
280                           & J0528.1-3933                     & LEDA17320                              & 0.037                          & 23.0                                                 & 18.1                                      & 1.7                              & Sy1                                 & single AGN iM                         & B                                \\
231                           & J0440.2-5941                     & LEDA15768                              & 0.058                          & 16.2                                                 & 18.2                                      & 3.2                              & Sy2                                 & single AGN iM                         & B                                \\
1627                          & J2328.1+0883                     & NGC7674                                & 0.028                          & 33.4                                                 & 18.8                                      & 4.0                              & Sy2                                 & single AGN iM                         & C                                \\
343                           & J0640.0-4737                     & SWIFTJ064013.50-474132.9               & 0.057                          & 20.3                                                 & 22.8                                      & 1.9                              & Sy2                                 & single AGN iM                         & N                                \\
1077                          & J2028.5+2543                     & NGC6921                                & 0.014                          & 91.0                                                 & 26.2                                      & 3.1                              & Sy2                                 & single AGN iM                         & A                                \\
1077                          & J2028.5+2543                     & MCG+04-48-002                          & 0.013                          & 91.0                                                 & 26.2                                      & 3.2                              & Sy2                                 & single AGN iM                         & A                                \\
712                           & J1413.2-0312                     & NGC5506                                & 0.006                          & 223.8                                                & 28.3                                      & 2.0                              & Sy1.9                               & single AGN iM                         & A                                \\
606                           & J1217.2-2611                     & ESO505-30                              & 0.04                           & 37.7                                                 & 29.6                                      & 2.2                              & Sy2                                 & single AGN iM                         & B                                \\
\\ \hline
\multicolumn{10}{c}{\textbf{Minor Galaxy Mergers}}
\\ \hline
318                           & J0606.0-2755                     & 2MASXJ06054896-2754398                 & 0.09                           & 1.7                                                  & 2.9                                       & 9.0                              & Sy1                                 & single AGN iM                         & D                                \\
430                           & J0843.5+3551                     & CASG218                                & 0.054                          & 2.7                                                  & 2.9                                       & 9.1                              & Sy2                                 & single AGN iM                         & D                                \\
134                           & J0234.6-0848                     & NGC985                                 & 0.043                          & 3.8                                                  & 3.4                                       & 6.9                              & Sy1                                 & single AGN iM                         & D                                \\
189                           & J0342.0-2115                     & ESO548-81                              & 0.014                          & 13.4                                                 & 3.8                                       & 8.8                              & Sy1                                 & single AGN iM                         & C                                \\
305                           & J0548.4-4748                     & LEDA17883                              & 0.05                           & 4.6                                                  & 4.5                                       & 4.8                              & Sy2                                 & single AGN iM                         & D                                \\
543                           & J1126.7+3514                     & Mrk423                                 & 0.032                          & 9.1                                                  & 6.0                                       & 4.8                              & Sy2                                 & single AGN iM                         & C                                \\
557                           & J1139.0-2323                     & HE1136-2304                            & 0.027                          & 11.8                                                 & 6.4                                       & 36.5                             & Sy1.9                               & single AGN iM                         & B                                \\
669                           & J1321.2+0859                     & LEDA46599                              & 0.033                          & 25.0                                                 & 6.6                                       & 13.8                             & Sy1.9                               & single AGN iM                         & C                                \\
303                           & J0544.4+5909                     & 2MASXJ05442257+5907361                 & 0.068                          & 4.8                                                  & 6.6                                       & 7.6                              & Sy1.9                               & single AGN iM                         & C                                \\
63                            & J0114.4-5522                     & NGC454E                                & 0.012                          & 27.8                                                 & 6.9                                       & 11.2                             & Sy2                                 & single AGN iM                         & C                                \\
1130                          & J2156.2+1724                     & 2MASXJ21561518+1722525                 & 0.034                          & 10.9                                                 & 7.3                                       & 5.3                              & Sy1                                 & single AGN iM                         & A                                \\
1439                          & J1416.5-3671                     & PKS1413-36                             & 0.074                          & 5.6                                                  & 8.6                                       & 7.8                              & Sy2                                 & single AGN iM                         & B                                \\
28                            & J0042.9-2332                     & NGC235A                                & 0.022                          & 19.6                                                 & 8.9                                       & 5.3                              & Sy1.9                               & single AGN iM                         & B                                \\
159                           & J0256.4-3212                     & ESO417-6                               & 0.017                          & 28.0                                                 & 9.2                                       & 82.7                             & Sy2                                 & single AGN iM                         & B                                \\
218                           & J0423.5+0414                     & LEDA15023                              & 0.046                          & 10.2                                                 & 9.3                                       & 29.3                             & Sy2                                 & single AGN iM                         & B                                \\
89                            & J0140.6-5321                     & 2MASXJ01402676-5319389                 & 0.072                          & 6.8                                                  & 9.5                                       & 24.1                             & Sy1.9                               & single AGN iM                         & B                                \\
1063                          & J2007.0-3433                     & CTS28                                  & 0.025                          & 18.8                                                 & 9.5                                       & 14.0                             & Sy1                                 & single AGN iM                         & C                                \\
302                           & J0544.4-4328                     & 2MASXJ05440009-4325265                 & 0.044                          & 15.1                                                 & 13.1                                      & 6.6                              & Sy2                                 & single AGN iM                         & B                                \\
136                           & J0238.2-5213                     & ESO198-24                              & 0.045                          & 15.1                                                 & 13.5                                      & 20.8                             & Sy1                                 & single AGN iM                         & B                                \\
217                           & J0422.7-5611                     & Fairall302                             & 0.044                          & 18.0                                                 & 15.5                                      & 4.4                              & Sy2                                 & single AGN iM                         & A                                \\
465                           & J0934.7-2156                     & ESO565-19                              & 0.016                          & 52.2                                                 & 17.6                                      & 8.7                              & Sy2                                 & single AGN iM                         & A                                \\
329                           & J0623.9-6058                     & ESO121-28                              & 0.04                           & 24.1                                                 & 19.1                                      & 7.3                              & Sy2                                 & single AGN iM                         & B                                \\
316                           & J0602.2+2829                     & IRAS05589+2828                         & 0.033                          & 28.3                                                 & 19.2                                      & 9.7                              & Sy1                                 & single AGN iM                         & A                                \\
193                           & J0350.1-5019                     & ESO201-4                               & 0.036                          & 27.4                                                 & 20.6                                      & 4.6                              & Sy2                                 & single AGN iM                         & B                                \\
416                           & J0823.4-0457                     & Fairall272                             & 0.022                          & 44.8                                                 & 20.6                                      & 877.7                            & Sy2                                 & single AGN iM                         & B                                \\
1255                          & J0306.0-3902                     & NGC1217                                & 0.021                          & 57.3                                                 & 24.4                                      & 11.6                             & LINER                               & single AGN iM                         & A                                \\
73                            & J0123.9-5846                     & Fairall9                               & 0.046                          & 27.3                                                 & 24.6                                      & 14.6                             & Sy1                                 & single AGN iM                         & B                                \\
1182                          & J2303.3+0852                     & NGC7469                                & 0.016                          & 78.6                                                 & 26.0                                      & 6.2                              & Sy1                                 & single AGN iM                         & A                                \\
471                           & J0945.6-1420                     & NGC2992                                & 0.008                          & 175.8                                                & 27.6                                      & 5.6                              & Sy1.9                               & single AGN iM                         & B                                \\
246                           & J0454.6-4315                     & LEDA146662                             & 0.087                          & 16.8                                                 & 28.5                                      & 10.0                             & Sy1.9                               & single AGN iM                         & A                                \\
197                           & J0354.2+0250                     & HE0351+0240                            & 0.036                          & 13.1                                                 & 29.6                                      & 36.2                             & Sy1                                 & single AGN iM                         & B                                \\  
\\ \hline
\multicolumn{10}{c}{\textbf{Single Galaxies}}
\\ \hline
342                           & J0640.4-2554                     & ESO490-26                              & 0.025                          & S                                              & S                                   & S                          & Sy1                                 & single AGN                            & D                                \\
584                           & J1200.2-5350                     & LEDA38038                              & 0.028                          & S                                              & S                                   & S                          & Sy2                                 & single AGN                            & N                                \\
1210                          & J2359.3-6058                     & PKS2356-61                             & 0.096                          & S                                              & S                                   & S                          & Sy2                                 & single AGN                            & N                                \\
489                           & J1009.3-4250                     & ESO263-13                              & 0.034                          & S                                              & S                                   & S                          & Sy2                                 & single AGN                            & N                                \\
1390                          & J1127.6-2912                     & ESO439-G009                            & 0.023                          & S                                              & S                                   & S                          & Sy2                                 & single AGN                            & N                                \\
1426                          & J1334.1-3842                     & WISEAJ133359.07-382450.3               & 0.052                          & S                                              & S                                   & S                          & Unknown                             & single AGN                            & N                                \\
83                            & J0131.8-3307                     & ESO353-9                               & 0.016                          & S                                              & S                                   & S                          & Sy2                                 & single AGN                            & N                                \\
442                           & J0902.8-7414                     & 2MASXJ09034285-7414170                 & 0.091                          & S                                              & S                                   & S                          & Sy2                                 & single AGN                            & N                                 
\enddata
\tablecomments{The table is divided according to stellar mass ratio (M1/M2) using M1/M2=4 as the separation between major and minor galaxy mergers. Each subset is sorted by projected separation ($D_{12}$ kpc). Columns: (1) target BAT ID; (2) SWIFT name; (3) Counterpart Name; (4) redshift measured from the BASS survey; (5)-(6) projected separation between the two nuclei in arcseconds and kpc. "S" mean that the source does not have a companion; (7) Mass ratio between primary and secondary galaxy based on NIR emission; (8) AGN type based on optical spectroscopy \citep{2022ApJS..261....2K}; (9) Type of merging AGN, as explained in Section \ref{sec:intro}; (10) Merger Stage, as explained in Section \ref{sec: morphology}}
\end{deluxetable*}

%assuming that the sources are at the same redshift. 

%Distribution of projected nuclear separation in kpc, for the 64 galaxies involved in merging systems in our sample; 35/64 galaxies are late-stage mergers (D12 < 10 kpc)

\subsection{Optical Morphologies} \label{sec: morphology}

% in the r-band in their respective surveys. For galaxies without optical images, we look for a r-band image in aladin in order to asses their morhpology.

\begin{figure*}[htbp]
    \centering
    \begin{subfigure}{0.19\textwidth}
        \centering
        \includegraphics[width=\textwidth]{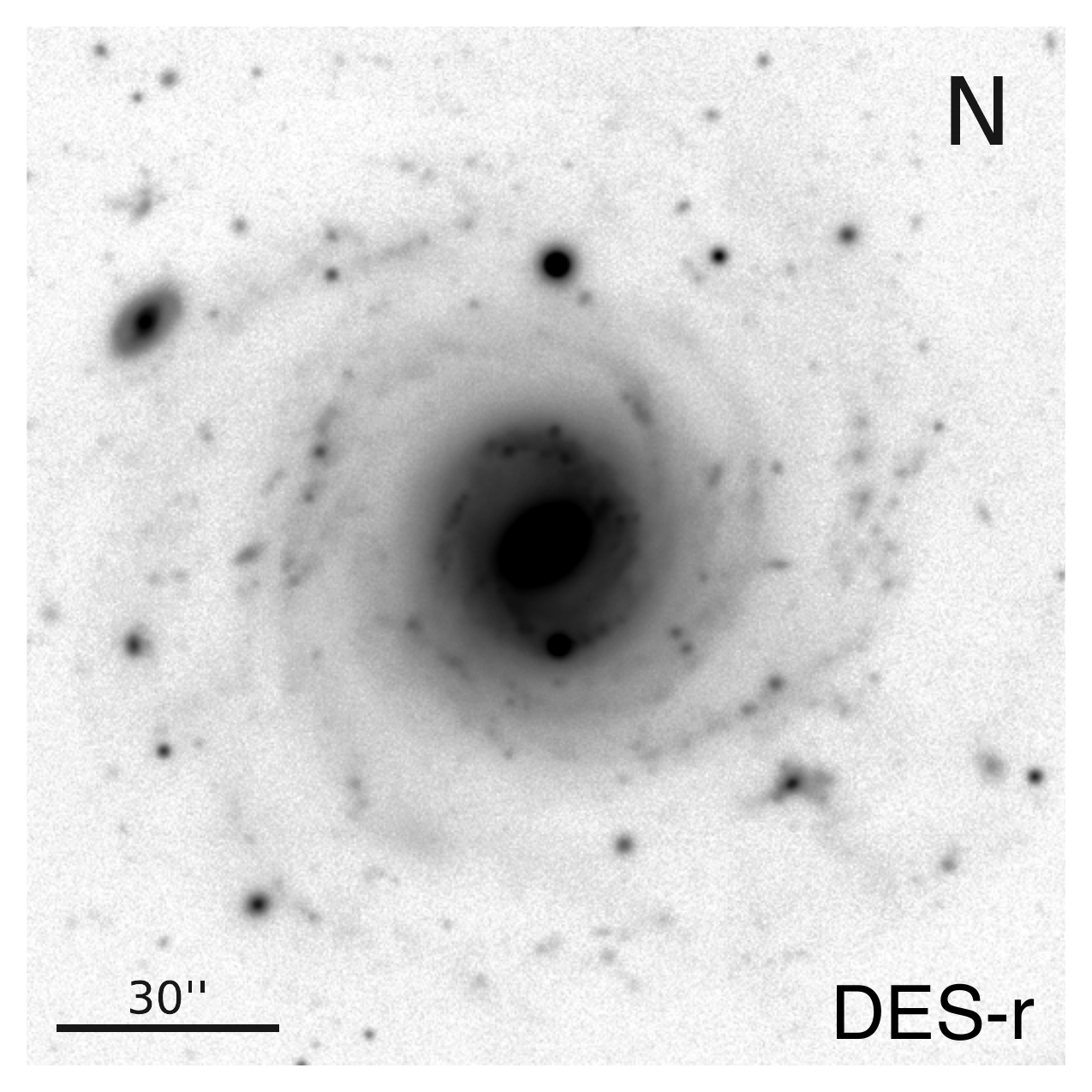}
    \end{subfigure}
    \begin{subfigure}{0.19\textwidth}
        \centering
        \includegraphics[width=\textwidth]{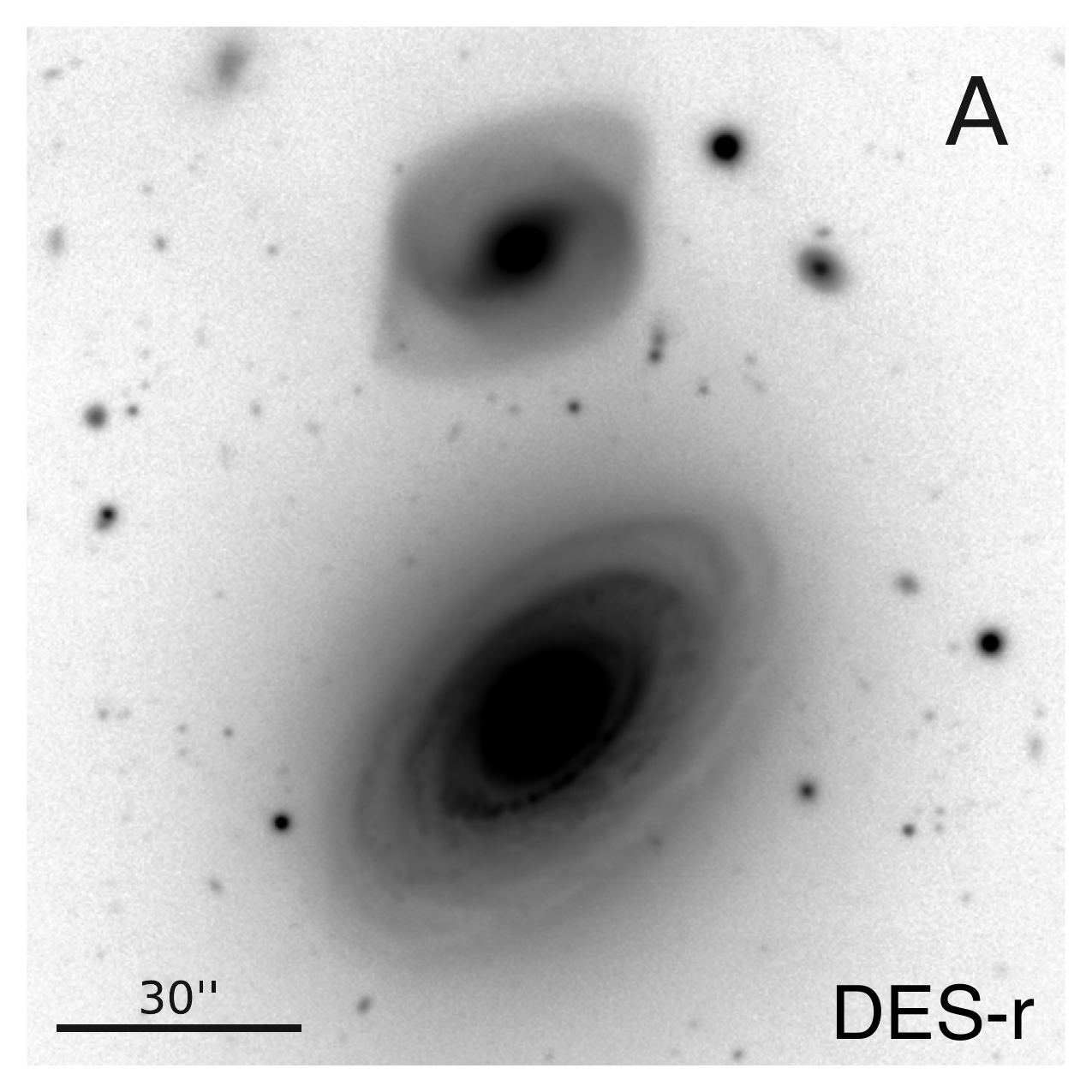}
    \end{subfigure}
    \begin{subfigure}{0.19\textwidth}
        \centering
        \includegraphics[width=\textwidth]{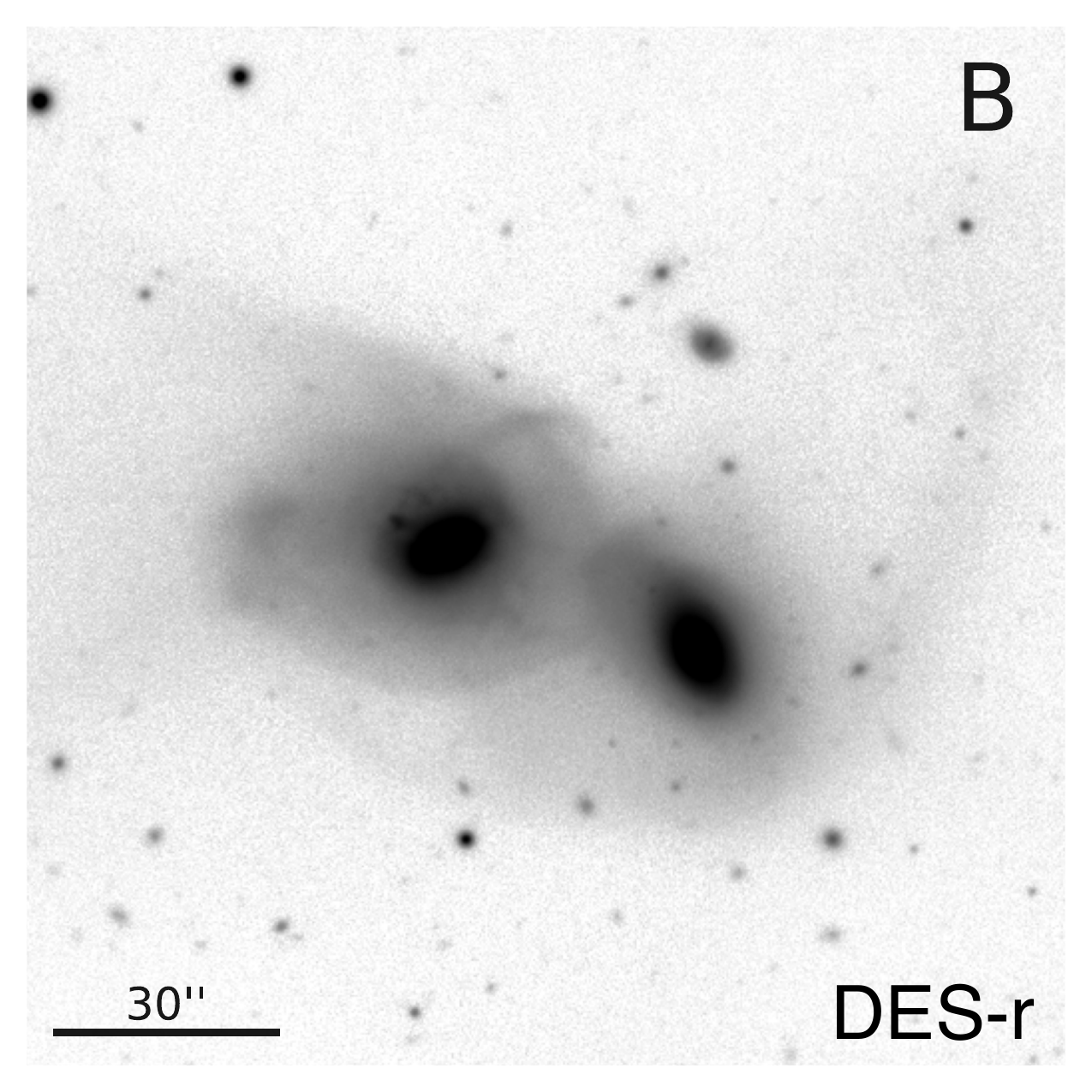}
    \end{subfigure}
    \begin{subfigure}{0.19\textwidth}
        \centering
        \includegraphics[width=\textwidth]{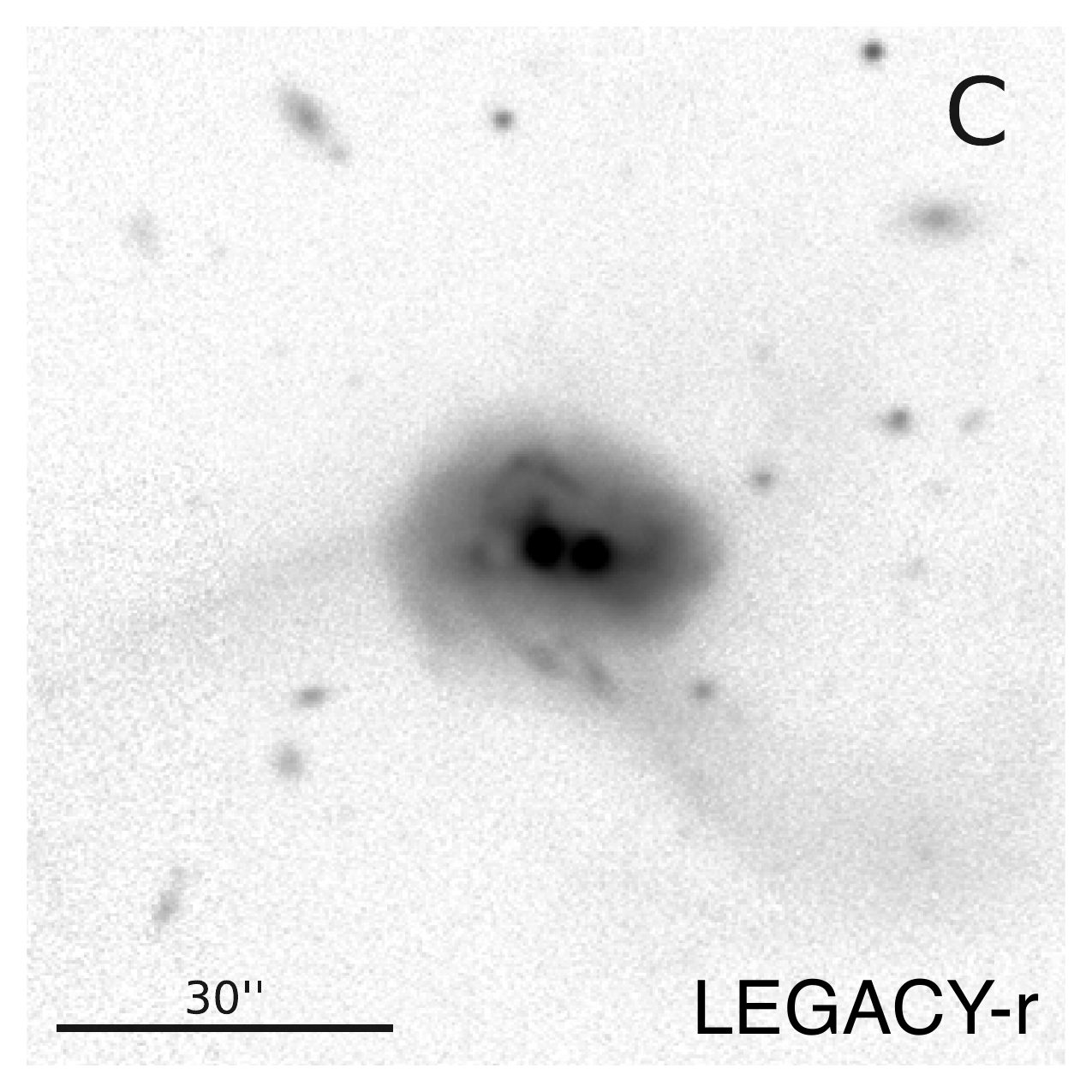}
    \end{subfigure}
    \begin{subfigure}{0.19\textwidth}
        \centering
        \includegraphics[width=\textwidth]{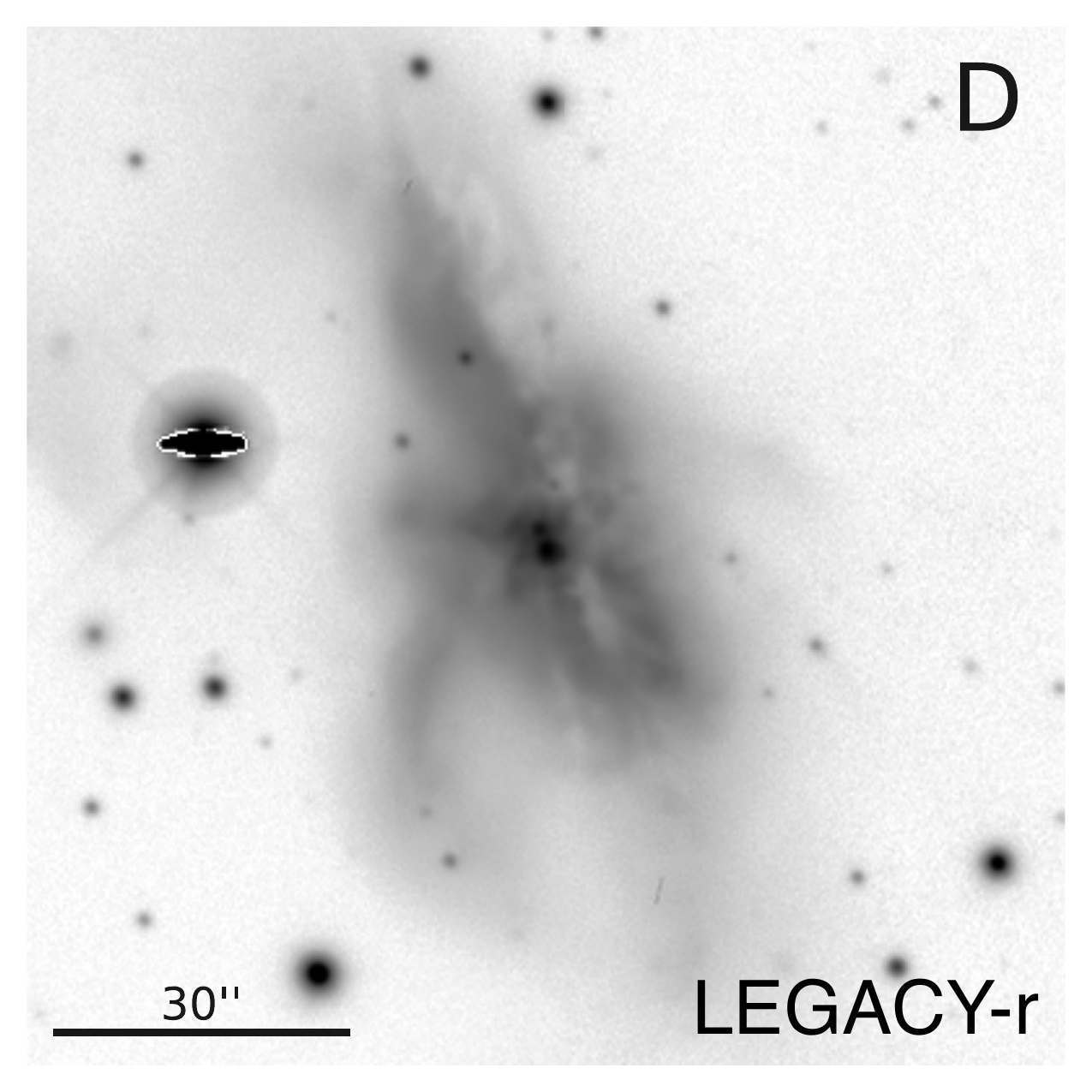}
    \end{subfigure}
    \caption{From left to right, N: Non-merging/isolated galaxies, A: pre-stage merger, B: early-stage merger, C: mid-stage merger, D: late-stage merger.} %Images in the r-band for its respective survey
    \label{fig:mer_st}
\end{figure*}

We visually classify our sources based on their optical morphology, using \textit{r}-band images from the available surveys for each object (e.g., DES, SDSS, DECaLS, PanSTARRS, etc.). To asses the merger stage, we examine characteristic merger features (such as tidal tails and/or disturbed disks) and the projected separation between the two nuclei, following the classification scheme of \cite{2013ApJS..206....1S}. 
The first stage, known as the \textit{pre-merger} (A), occurs when two galaxies begin to interact and move closer to one another, without physical contact. In the \textit{early-stage merger} (B), both galaxies have had their first encounter but still maintain symmetric disks, with only faint tidal features. The system evolves into a \textit{mid-stage merger} (C), characterized by amorphous disks and the most pronounced merger features, as the interaction continues. Finally, in the \textit{late-stage merger} (D), both nuclei merge into a single envelope, just before coalescence. Additionally, there is a category for non-merging systems (N), in which we include sources that do not show evidence of two distinct nuclei. Morphological data alone does not allow us to distinguish if these cases correspond to isolated non-merging systems (i.e., N comes before A) and post-merger systems (i.e., N comes after D). Likely, this bin includes a mix of them. Figure \ref{fig:mer_st} presents an example of a system in each stage. While human-based morphological classification is inherently subjective, any ambiguities in assigning merger stages, aside from the N category, will not affect the overall trends and general conclusions presented in this study.

\subsection{Control Sample}

To compare our merging sample with non-interacting galaxies, we select 532 non-active galaxies from xCOLD GASS with the measurements of SFR and stellar mass ($M_*$) from \citet{2017ApJS..233...22S}. These galaxies lie in the redshift range $0.01 < z < 0.05$, with SFRs between $0.02<\mathrm{SFR}/M_{\odot}\ \mathrm{yr}^{-1}<34.5$ and stellar masses $M_*$$>$10$^9\ M_\odot$.

Moreover, we include 213 local AGN-host galaxies (0.01$<$$z$$<$0.05) from the BASS sample with available measurements of SFR ($0.02<\mathrm{SFR}/M_{\odot}\ \mathrm{yr}^{-1}<132$) and stellar mass ($10^{9.5}<M_*/M_{\odot}<10^{11.5}$), provided by \citet{2021ApJS..252...29K}, who also compare these AGN hosts directly with the xCOLD GASS galaxies. Furthermore, these sources have measurements of AGN bolometric luminosities ($10^{43}<L_{bol}/\mathrm{erg\ s^{-1}}<10^{46}$), as reported by \citet{2017ApJS..233...17R}. 

\section{Photometry} \label{sec: photometry}

We use all available photometric bands from the X-ray to the FIR to cover a wide wavelength range for SED fitting. To ensure accurate relative photometric calibrations, we developed a set of custom procedures in Python, described in Section \ref{sec: fluxes}, to match the apertures and resolutions of the sources in the different photometric bands, in order to minimize wavelength-dependent differences in the fraction of the integrated emission, addressing inconsistencies in aperture sizes and measurement methods in the available UV-to-FIR survey data, which can significantly impact SED fitting. This is done by incorporating images from different surveys. At X-ray wavelengths, we use the intrinsic fluxes obtained from \cite{2017ApJS..233...17R}, as will be presented in Section \ref{sec:xrays}. To complete our photometric study, we incorporate measurements from the following surveys:

\begin{itemize}[leftmargin=12pt]
    \item \textbf{UV}: data obtained from the Galaxy Evolution Explorer (GALEX; \citealp{2005ApJ...619L...1M}) in its Far-UV (153 $\mu$m) and Near-UV (229 $\mu$m) bands (spatial resolutions of 4.5" and 6.0" respectively).
    \item \textbf{Optical}: We use optical data from three surveys to ensure coverage for all our sources in the entire sky: (1) The Dark Energy Survey (DES; \citealp{2018ApJS..239...18A}) in the \textit{grizY} filters (0.48, 0.62, 0.75, 0.87, and 1.02~$\mu$m;  spatial resolution from 0.9" to 1.2"), (2) Sloan Digital Sky Survey (SDSS; \citealp{2000AJ....120.1579Y}) in the \textit{ugriz} filters (0.36, 0.47, 0.62, 0.75, and 0.89~$\mu$m;  spatial resolutions ranging from 1.3" to 2.0") and, (3) The DECam Legacy Survey (DECaLS; \citealp{2019AJ....157..168D}) in the \textit{griz} filters (0.48, 0.62, 0.75, and 0.87~$\mu$m;  spatial resolutions from 1.0" to 1.2").
    \item \textbf{NIR}: For the near-IR emission, we use the Two Micron All Sky Survey (2MASS; \citealp{2006AJ....131.1163S}) in the \textit{J} (1.25~$\mu$m), \textit{H} (1.65~$\mu$m), and \textit{Ks} (2.17~$\mu$m) filters (spatial resolutions from 2.0" to 2.5").
    \item \textbf{MIR}: Mid-IR data are taken from the Wide-field IR Survey Explorer (WISE; \citealp{2010AJ....140.1868W}) using the AllWISE data release, in its W1(3.4~$\mu$m), W2 (4.6~$\mu$m), W3 (12~$\mu$m), and W4 (22~$\mu$m) (spatial resolution from 6.1" to 12.0").
    \item \textbf{FIR}: For far-IR emission, we use the Photodetector Array Camera and Spectrometer (PACS; \citealp{2010A&A...518L...2P}) onboard Herschel in its $70\ \mu m$ and $160\ \mu m$ bands (spatial resolutions of 5.8" and 12.0" respectively).
\end{itemize}

Not all sources have available data in all bands. Table \ref{tab:summary_bands} summarizes the information about the availability of bands for each source. One source, BAT ID 343, was excluded from the final sample because its fluxes could not be measured due to the presence of a bright nearby star, which affected the measurements and prevented CIGALE from obtaining a reliable fit. As a result, the final sample consists of 71 sources.

\newlength{\mytabcolsep}
\setlength{\mytabcolsep}{1.5pt} % Ajusta este valor según lo necesario

\begin{table}[htbp]
\centering
\setlength{\tabcolsep}{\mytabcolsep} % Ajusta el espacio entre columnas solo para esta tabla
\begin{tabular}{c|c|c|c|c|c|c}
%\toprule
X-RAYS & UV & OPTICAL & NIR & MIR & FIR & N° sources \\ \midrule
\checkmark & \checkmark & \checkmark & \checkmark & \checkmark & \checkmark & 25         \\
\checkmark & \checkmark & \checkmark & \checkmark & \checkmark & x    & 19        \\
\checkmark & \checkmark & x & \checkmark & \checkmark    & \checkmark & 4        \\
\checkmark & x & \checkmark & \checkmark & \checkmark    & \checkmark & 1        \\
x & \checkmark & \checkmark & \checkmark & \checkmark    & \checkmark & 1        \\       
\checkmark & \checkmark & \checkmark     & x & \checkmark    & \checkmark & 1         \\
\checkmark & \checkmark     & x & \checkmark & \checkmark & x & 2         \\
\checkmark & x     & \checkmark & \checkmark & \checkmark & x & 5         \\
\checkmark & x     & x & \checkmark     & \checkmark & \checkmark & 3         \\
x & \checkmark     & \checkmark & \checkmark & \checkmark & x & 5         \\
\checkmark & x & x & \checkmark & \checkmark & x    & 1         \\
x & \checkmark & x & \checkmark & \checkmark    & x & 2         \\
x & x & x     & \checkmark & \checkmark & \checkmark & 2         \\
\hline
\multicolumn{6}{l}{\textbf{Total}} & 71\\
\bottomrule
\end{tabular}
\caption{Available multiwavelength photometry used on the SED fitting. }
\label{tab:summary_bands}
\end{table}

\subsection{Flux Measurements} \label{sec: fluxes}

By selection, the sources in our sample exhibit complex and disturbed morphologies, making it challenging to obtain consistent apertures and resolutions across different surveys. To address this, we follow the procedure developed by Rojas et al. (in prep.) to get reliable photometric measurements while minimizing contamination from nearby objects and ensuring that the bands' apertures and spatial resolutions are matched. In the following, we outline the steps in our methodology:

\begin{itemize}[leftmargin=12pt]
    \item \textbf{Identifying target galaxies and surrounding sources}: To identify well-detected sources, we first apply a 2-sigma threshold above the background to identify significant emission. Then, using the system coordinates, we locate the source of interest. If, at this threshold, we do not identify low-brightness features, which are essential for the photometry, we manually define a precise region using DS9 for each band. This region is also helpful for excluding unrelated nearby sources that are not part of the galaxy. This step ensures that the primary galaxy is accurately identified while minimizing contamination from adjacent sources. If the merging system is resolved in the lowest-resolution band (W4), we target only the BASS source (7/72). However, if it remains unresolved in W4, we include both the BASS source and its counterpart to ensure consistency across all wavelengths (65/72).
    \item \textbf{Surrounding sources removal/replacement}: Once surrounding sources are identified (either automatically or manually via DS9 regions), they are removed by replacing their pixel values using random background values, from the entire field, to simulate an image containing only the target surrounded by background noise. If a foreground star overlaps with the galaxy, it is identified using the GAIA DR3 \citep{2023A&A...674A...1G} catalog and subsequently removed while extrapolating the galaxy's brightness profile in the region occupied by the star. This procedure introduces a small increase in the local deviation within the replaced region compared to the rest of the image, particularly in similar regions. However, this effect is smoothed out in the subsequent processing steps.
    \item \textbf{Resolution matching}: To ensure consistent resolution across all the bands, we convolve all images to achieve a final angular resolution of $\sim 12"$, which matches the resolution of the WISE W4 band. This step is critical for minimizing systematic biases introduced by differences in the spatial resolution of the data. 
    \item \textbf{Flux measurements}: Finally, after the convolution, fluxes are measured using a single aperture across all bands. Each source's aperture size is individually chosen based on the morphology and inclination, ensuring that it is large enough to encompass the whole galaxy and captures all of its flux in all the available bands, while remaining as small as possible. To achieve this, we adopt a human-guided aperture selection method, using flux-aperture plots to verify that the chosen aperture includes the galaxy's total emission. In addition, an annular region is used to estimate the local background. This uniform approach enables consistent and reliable measurements across the entire spectral range.
\end{itemize}

\begin{figure*}[htbp]
    \centering
    \includegraphics[width=\textwidth]{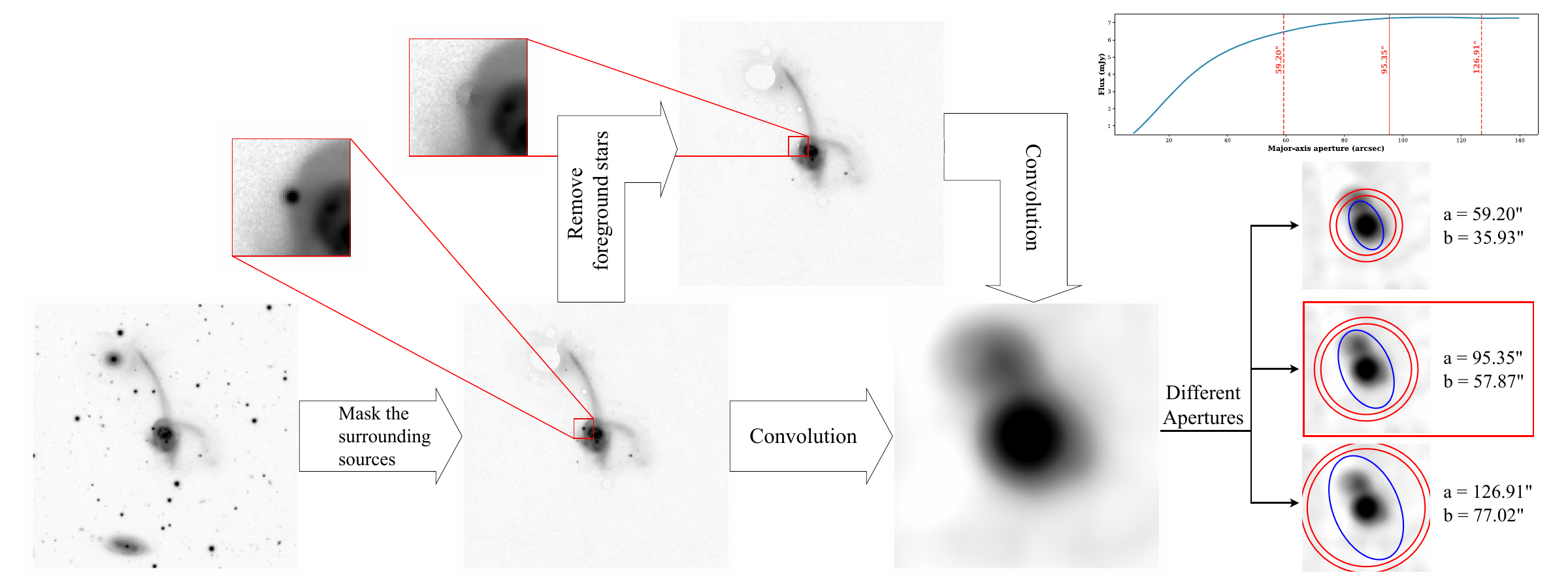}
    \caption{Schematic visualization of the custom-made code developed for flux measurements. From left to right: the first panel shows the stacked image downloaded from the relevant survey (in this case, as an example, we show the source BAT ID 305 in the DES-r band). The second step removes the surrounding sources, replacing their pixels with random values obtained from the background. In the third step, if a foreground star lies along the line of sight, it is removed by extrapolating the galaxy's brightness profile across the region occupied by the star. If no foreground star is present, this step is skipped. In the fourth panel, we convolve the image to reduce its resolution to match the resolution of the WISE 4 band, which has the lowest. Finally, multiple apertures are applied to determine the optimal one for consistent measurements across all available bands (blue ellipses with "a" and "b" the ellipse axes). A circular annulus is also used to estimate the local background (red circles). The optimal aperture is selected as the smallest that fully encompasses the entire galaxy, guided by the flux-aperture plot. In this example, the chosen aperture is indicated by a red rectangle and the solid red line in the flux-aperture plot.} 
    \label{fig:code}
\end{figure*}

Figure \ref{fig:code} presents a graphical representation and an example of the photometric method followed here.

\subsection{Photometric Errors}

To estimate the photometric uncertainties, we consider two main sources of random noise: Poisson fluctuations from the source signal and background noise from the surrounding sky.

The Poisson noise is estimated from the square root of the source counts, while the background noise is measured from the standard deviation of the background values in an annular region around the source. These components are combined and scaled by a conversion factor ($k$), which transforms counts (or instrumental units) into physical flux units (mJy). The total uncertainty is then computed as:

\begin{equation}
   \sigma_{\mathrm{mJy}} = k\sqrt{(\sqrt{\mathrm{counts}})^2 + \sigma_{\mathrm{bkg}}^2}
\end{equation}

For surveys where the detector gain is available, such as 2MASS and DES, we use the \texttt{calc\_total\_error}\footnote{\url{https://photutils.readthedocs.io/en/stable/api/photutils.utils.calc_total_error.html}} function from the \texttt{photutils} package. This method provides a more accurate estimate of the total error by properly incorporating gain information, read noise, and background variance.

\subsection{Foreground Galactic Extinction Correction} \label{sec:exctintion}

The fluxes from the FUV to W2 bands were corrected for galactic extinction to account for the Milky Way reddening. For this, we utilized the Python library \texttt{dustmaps} \citep{2018JOSS....3..695M}, which enabled us to determine the reddening value $E(B-V)_{SFD}$ for each source. The individual extinction values were then calculated using the following relation:

\begin{equation}
    A_{\lambda} = R_{\lambda}E(B-V)_{SFD},
\end{equation}

\noindent
where $R_{\lambda}$ is the extinction coefficient for each band. These values were obtained from \cite{2011ApJ...737..103S} for SDSS, DES, and DECaLS, and from \cite{2013MNRAS.430.2188Y} for GALEX, 2MASS, and WISE. This correction was negligible for bands redder than W2 and hence was not applied in those cases.

\subsection{X-ray fluxes} \label{sec:xrays}

Our photometric measurements were essential for recovering the total emission from each source. In contrast, X-ray emission originates from compact regions in the nucleus of a galaxy, making flux losses negligible. Therefore, we can directly use previous measurements of the X-ray fluxes. All galaxies in our study belong to the Swift/BAT sample and, hence, have hard X-ray measurements available in the literature. Specifically, we use the intrinsic (i.e., obscuration-corrected) X-ray fluxes reported by \citet{2017ApJS..233...17R} for the 2--10 keV and 14--195 keV bands. These fluxes were derived through spectral analysis using data from XMM-Newton, Swift/XRT, ASCA, Chandra, and Suzaku for soft X-rays and from Swift/BAT for hard X-rays. The resulting parameters include estimates of column densities ($N_H$) and measurements of the photon index ($\Gamma$), which is then given to \textsc{CIGALE} as input values.

As a caveat, we remark that there is, on average, a spread of up to $\sim$10 years between the X-ray observations and the remaining multi-wavelength datasets (with 2MASS near-IR photometry showing the largest offsets). Consequently, some of the sources in our study could be affected by  AGN variability. Nevertheless, since Swift/BAT fluxes are averaged over the years, short-term X-ray variability is significantly smoothed out.

\section{SED modeling} \label{sec: sed}

To constrain the physical properties of our sample, we employed the Code Investigating GALaxy Emission (CIGALE; \citealp{2019A&A...622A.103B}) to perform SED fitting for each source. The fitting utilizes the photometry obtained in Section \ref{sec: photometry}, which spans the UV to FIR range, and the X-ray fluxes from Section \ref{sec:xrays}.

The host galaxy component was modeled using a delayed star formation history with an additional burst episode, as we expect galaxy mergers to undergo a starburst event. The stellar emission was represented by a single stellar population, as modeled by \cite{2003MNRAS.344.1000B}. For dust attenuation, we adopted a modified version of \cite{2000ApJ...533..682C}. Dust emission was modeled using the templates from \cite{2014ApJ...780..172D}. The AGN emission was characterized using the SKIRTOR templates \citep{2012BlgAJ..18c...3S}, which account for the clumpy structure of the AGN torus. In particular, we modeled two AGN components using a modified version of CIGALE for the confirmed dual AGNs in our sample. The two components differ primarily in the AGN fraction parameter: in the first AGN component, this fraction represents the ratio of the AGN IR emission to the total IR emission from the merging system, whereas, for the second component, it represents the contribution of the second AGN instead of the first. Finally, CIGALE utilizes a specific module, as described by \cite{2020MNRAS.491..740Y}, to model the X-ray emission. As mentioned in Section \ref{sec:xrays}, the photon index was provided as input to CIGALE, ensuring that the information from the hard X-ray band was incorporated. The parameter space and corresponding input values used for the SED fitting are summarized in Table \ref{tab:sed modules}. 

\begin{table*}[htbp]
\centering
\begin{tabular}{llll}
\hline
Model            & Module                       & Parameter                 & Value                                       \\ \hline
SFH model        & \texttt{sfhdelayedbq}        & \texttt{tau\_main}                 & 1000, 3000, 5000                            \\
                 &                              & \texttt{age\_main}                 & 4500, 7000, 9500, 12000                     \\ 
                 &                              & \texttt{age\_bq}                   & 10, 20, 100                                 \\
                 &                              & \texttt{r\_sfr}                    & 0.1, 1, 10, 1000                            \\ \hline
Stellar emission & \texttt{bco3}                         & \texttt{imf}                       & 1                                           \\ \hline
Attenuation law  & \texttt{dustatt\_modified\_starburst} & \texttt{E\_BV\_lines}              & 0.1, 0.5, 1.0, 1.5, 2.0                     \\
                 &                              & \texttt{powerlaw\_slope}           & -0.8, -0.4, 0.0                             \\ \hline
Dust emission    & \texttt{dl2014}                       & \texttt{qpah}                      & 0.47, 1.12, 2.50                            \\
                 &                              & \texttt{umin}                      & 1, 5, 10, 25, 50                            \\
                 &                              & \texttt{alpha}                     & 2.0, 2.5                                    \\ \hline
AGN              & \texttt{skirtor2016}                     & \texttt{oa}                        & 20, 40, 60, 80                              \\
                 &                              & \texttt{i}                         & 0, 30, 60, 90                               \\
                 &                              & \texttt{fracAGN}                   & 0.1, 0.2, 0.3, 0.4, 0.5, 0.6, 0.7, 0.8, 0.9 \\
                 &                              & \texttt{EBV}                       & 0.0, 0.05, 0.3, 0.8                         \\ 
                 & \texttt{skirtor2016\_2\footnote{This model is only used for dual AGNs.}}                     & \texttt{oa}                        & 20, 40, 60, 80                              \\
                 &                              & \texttt{i}                         & 0, 30, 60, 90                               \\
                 &                              & \texttt{fracAGN}                   & 0.1, 0.2, 0.3, 0.4, 0.5, 0.6, 0.7, 0.8, 0.9 \\
                 &                              & \texttt{EBV}                       & 0.0, 0.05, 0.3, 0.8                         \\ \hline
X-rays           & \texttt{xray}                       & \texttt{gam}                     & 1.4, 1.8, 2.2, 2.6                      \\
                 &                              & \texttt{alpha\_ox}                     & -1.9, -1.7, -1.5, -1.3, -1.1                 \\ \hline
\end{tabular}
\caption{CIGALE input parameters, generating $7.46\times10^9$ models. For parameters not listed here, default values have been adopted.}
\label{tab:sed modules}
\end{table*}

To manage the increased complexity introduced by modeling two AGN components, we adopted a two-step fitting approach. In the first step, we fit our 71 sources using a single AGN component to constrain all parameters, generating approximately $7.46\times10^9$ models. In the second step, we reran the code for the confirmed nine dual AGNs, now fixing the galaxy parameters obtained in the first step while allowing the two AGN components to vary independently, generating approximately $3.6\times10^5$ models. This strategy significantly reduces computational time, enabling a more accurate characterization of individual AGN parameters. The impact of using one or two AGN components is discussed in Section \ref{sec:impact}.

After running the fitting procedure, source BAT ID 497 failed to converge to a good solution and was therefore excluded from the final sample, resulting in 70 sources with available measured parameters.

To ensure the reliability of our parameter estimations, we utilized the mock analysis tool provided in CIGALE to validate our results. This tool generates synthetic observations based on the best-fit models and compares them to the input data, allowing us to assess the robustness of the derived physical parameters. Overall, the values for the mock galaxies are consistent with those derived from the observed data, obtaining similar physical parameters. This consistency indicates that our results are not significantly affected by degeneracies. Further details about this procedure are provided in Appendix \ref{app: mock}.

\section{Results} \label{sec: results}

%In this work, we focus on the relations between the AGN emission and the SFR of the host galaxy, that were measured using a multiwavelength SED fitting approach from X-rays to FIR for a sample of 63 merging systems and 7 single galaxies from the BASS sample.

\begin{figure}[htbp]
    \centering
    \includegraphics[width=\columnwidth]{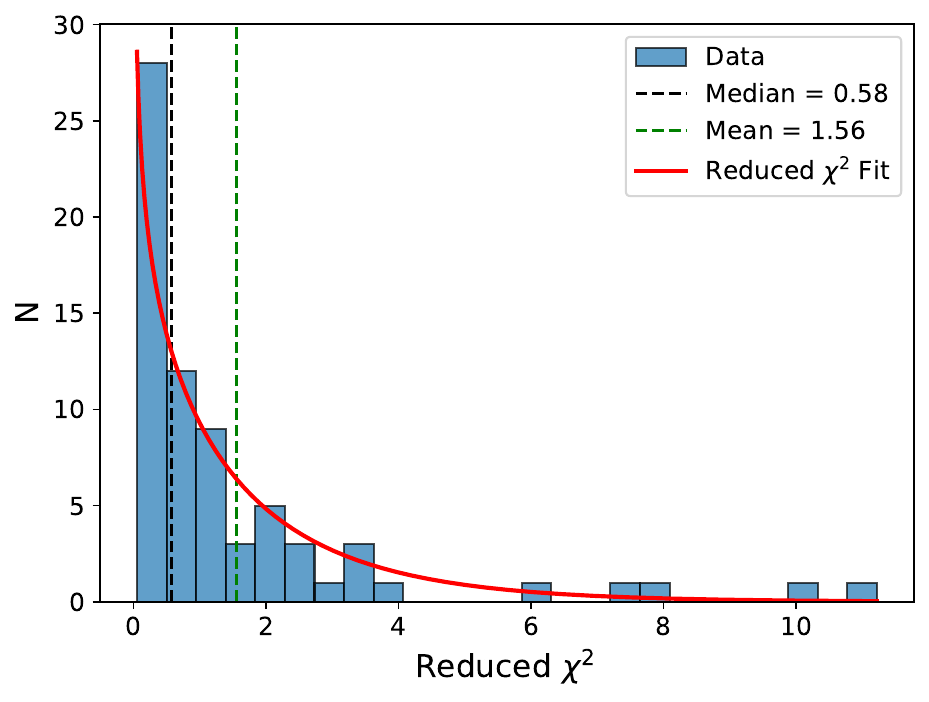}
    \caption{Reduced $\chi^2$ distribution for the fits performed by CIGALE. The red line represents the theoretical $\chi^2$ function for our degrees of freedom ($dof$), demonstrating that our results are consistent with expectations. Furthermore, 87\% of our fits have $\chi^2<3$ values.}
    \label{fig:chi_2}
\end{figure}

In this section, we present the physical properties of our sample, derived using CIGALE, as a function of the projected separation in their merging event. Figure \ref{fig:chi_2} shows the reduced $\chi^2$ distribution for our fits, with a mean of 1.56 and a median of 0.58, alongside the fitted $\chi^2$ distribution, closely matching our results. A total of 87\% of our fits have reduced $\chi^2<3$, indicating a high fraction of well-modeled sources. However, our mean value is higher than the expected $\chi^2$ = 1, due to the presence of a small fraction of poorly fitted sources.

%and that's supported by our median value of $\chi^2=0.6<1$

\begin{figure*}[htbp]
    \centering
    \begin{subfigure}{0.49\textwidth}
        \centering
        \includegraphics[width=\textwidth]{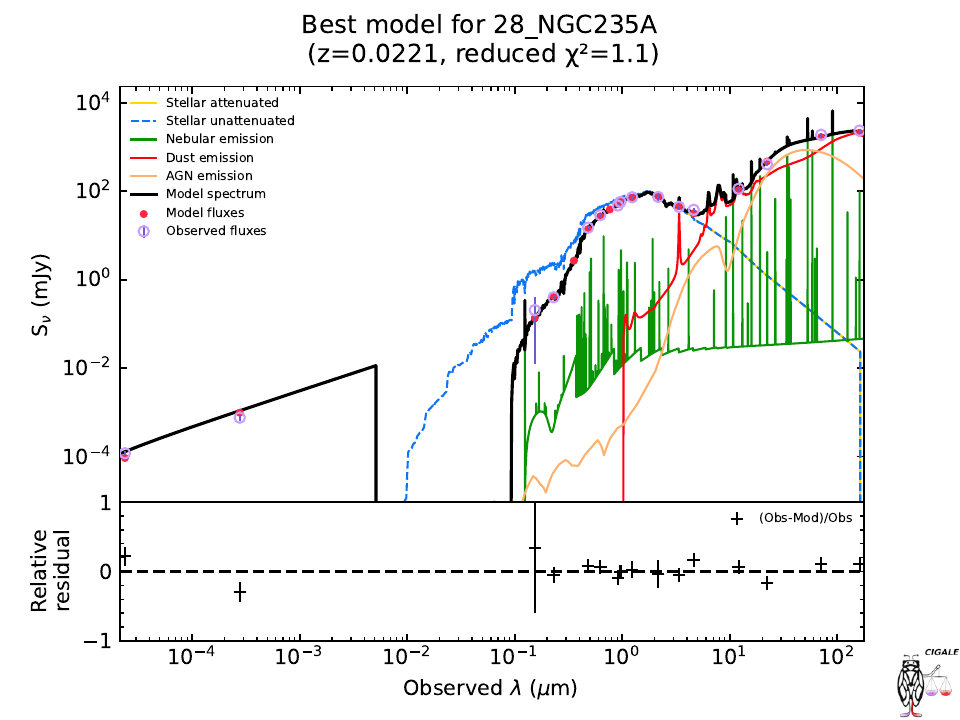}
    \end{subfigure}
    \begin{subfigure}{0.49\textwidth}
        \centering
        \includegraphics[width=\textwidth]{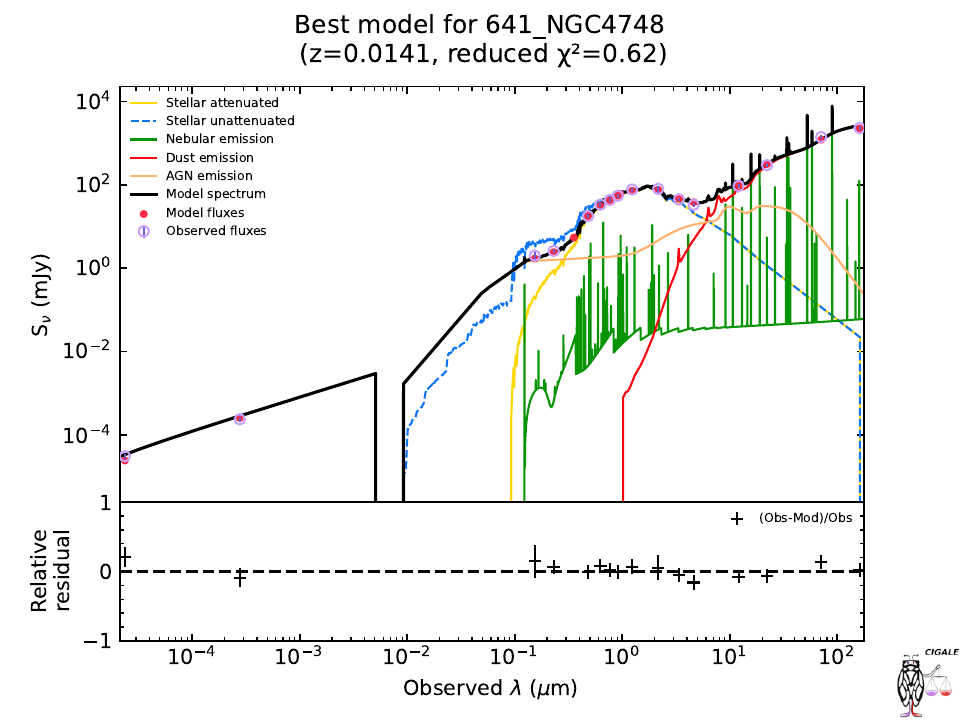}
    \end{subfigure}
    \caption{Best-fit SEDs for BAT ID 28 and BAT ID 641 obtained using CIGALE. Blue circles represent the observed fluxes calculated in Section \ref{sec: fluxes}, while the red points are the model-predicted fluxes. The reduced $\chi^2$ value quantifies the quality of each fit, and the residuals are shown in the lower panel of each subplot. The complete figure set (70 images) is available in the online journal.}
    \label{fig: sed_examples}
\end{figure*}

Although some sources exhibit higher $\chi^2$ values, the fraction of $\sim\ 10\%$ bad fits is consistent with expectations from the theoretical reduced $\chi^2$ distribution. Some poorly fitted sources ($3<\chi^2<6$) are expected based on statistical predictions, and their absence would be unlikely. Our $\chi^2$ distribution predicts $\sim$10 sources in this range, while we observe 5. However, four sources display very large $\chi^2$ values ($>6$), which is unexpected. They should instead fall within the $3<\chi^2<6$ range for full consistency with the predicted distribution. The reasons for these poor fits are further discussed in Section \ref{sec:impact}. Nevertheless, our overall results closely match the theoretical distribution, which reinforces our confidence in the modeling approach employed here. The discrepancy is limited to only four sources, which is statistically negligible. Additionally, we validate our results using the mock analysis provided by CIGALE, as mentioned in Section \ref{sec: sed}. One best-fit model is presented in Figure \ref{fig: sed_examples} and all the sed models are in 

\figsetstart
\figsetnum{6}
\figsettitle{CIGALE Best SED Models for All Sources}

\figsetgrpstart
\figsetgrpnum{6.1}
\figsetgrptitle{13}
\figsetplot{seds/13_LEDA136991_best_model.pdf}
\figsetgrpnote{Spectral Energy Distribution for BAT ID: 13}
\figsetgrpend

\figsetgrpstart
\figsetgrpnum{6.2}
\figsetgrptitle{28}
\figsetplot{seds/28_NGC235A_best_model.pdf}
\figsetgrpnote{Spectral Energy Distribution for BAT ID: 28}
\figsetgrpend

\figsetgrpstart
\figsetgrpnum{6.47}
\figsetgrptitle{641}
\figsetplot{seds/641_NGC4748_best_model.pdf}
\figsetgrpnote{Spectral Energy Distribution for BAT ID: 641}
\figsetgrpend

\figsetend

\begin{figure}[htbp]
    \centering
    \includegraphics[width=\columnwidth]{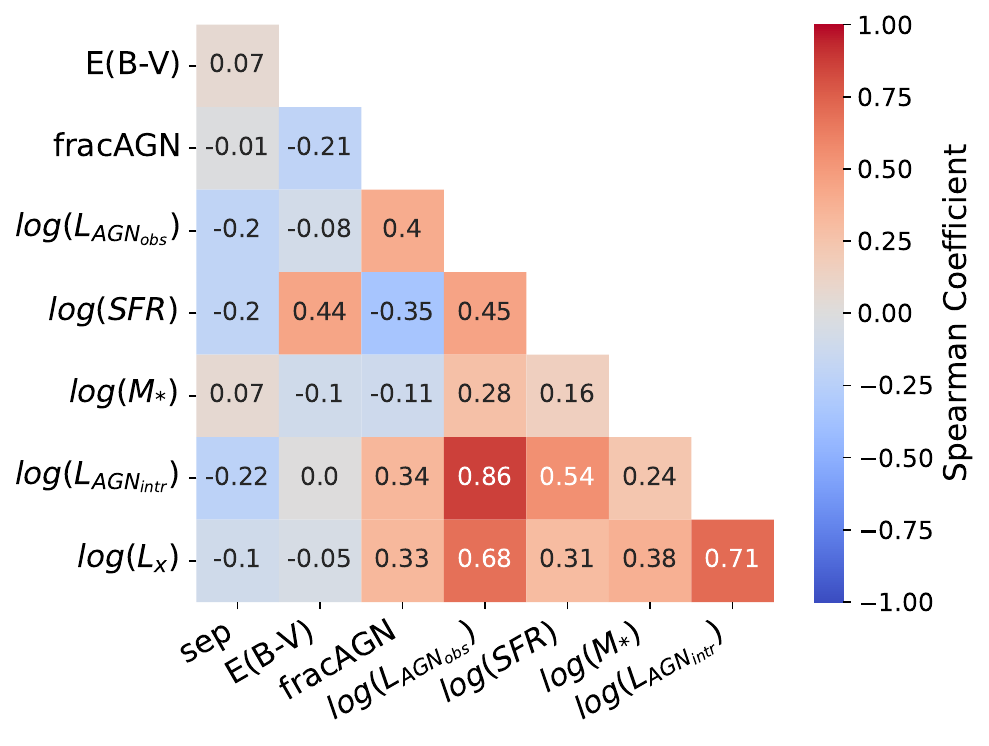}
    \caption{Confusion matrix of Spearman correlation values for pairs of physical parameters. Blue squares indicate a negative (anti-correlated) relationship, while red squares indicate a positive (correlated) relationship between the parameters.} 
    \label{fig:spearman}
\end{figure}

Figure \ref{fig:spearman} shows the degree of correlation among the most important measured parameters, where redder squares indicate stronger correlations, bluer ones show stronger anti-correlation, and white represents no correlation. Distinct patterns can be observed, such as the absence of a clear correlation between the SFR and $M_*$, where the so-called star formation main sequence of galaxies might be expected for isolated sources. However, we do not find evidence of a relationship in our merger sample. In Section \ref{sec: MS}, we discuss this topic in depth. Additionally, we identify a slight anti-correlation between the SFR and the AGN IR luminosity fraction, suggesting possible effects from feedback caused by the more luminous AGN. Alternatively, this trend may also arise from the model's IR emission distribution, where AGN and star formation contribute, making them appear linked. Nevertheless, a mild correlation exists between the intrinsic AGN luminosity and the SFR, indicating a potential co-evolution between the SMBH and the host galaxy during the merging process. Both topics are further discussed in Section \ref{sec: AGN}.

\subsection{Correlations for Major Mergers}
\label{sec: major_coeffs}

\begin{figure}[htbp]
    \centering
    \includegraphics[width=\columnwidth]{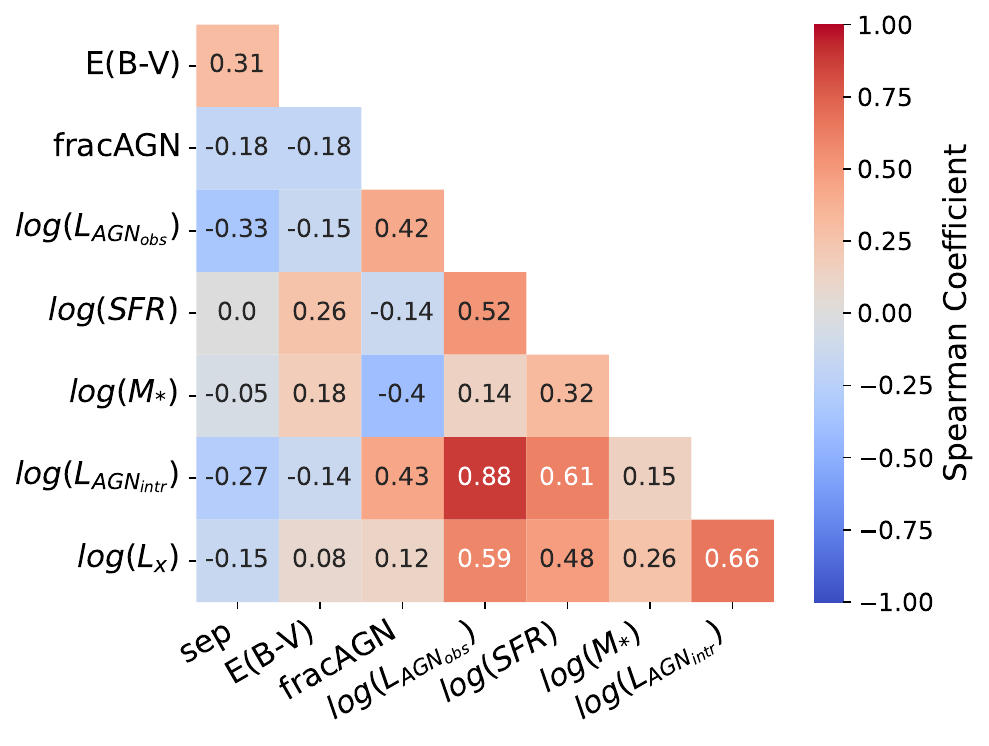}
    \caption{Confusion matrix of Spearman correlation values for pairs of physical parameters, considering only systems classified as major galaxy mergers. Colors are same as Figure \ref{fig:spearman}.} 
    \label{fig:spearman_major}
\end{figure}

We further investigate correlations between physical parameters by separating major and minor galaxy mergers, using a mass ratio of 4:1 as the dividing boundary. Figure \ref{fig:spearman_major} presents the correlation matrix for the physical parameters derived from our CIGALE fits, specifically considering only major galaxy mergers. Overall, there are no significant differences with the full sample, thus indicating that there are no major biases in these two samples. The only exception is the correlation between SFR and AGN luminosity, which we find is appreciably weaker in minor mergers (Spearman coefficient of 0.38), compared to the whole sample. Consequently, we can conclude that the link between star formation and AGN activity is more significant for major galaxy mergers.

%(M1/M2 $<$ 4)

\subsection{Separation} \label{sec: sep}

\begin{figure*}[htbp]
    \centering
    \begin{subfigure}{0.49\textwidth}
        \centering
        \includegraphics[width=\textwidth]{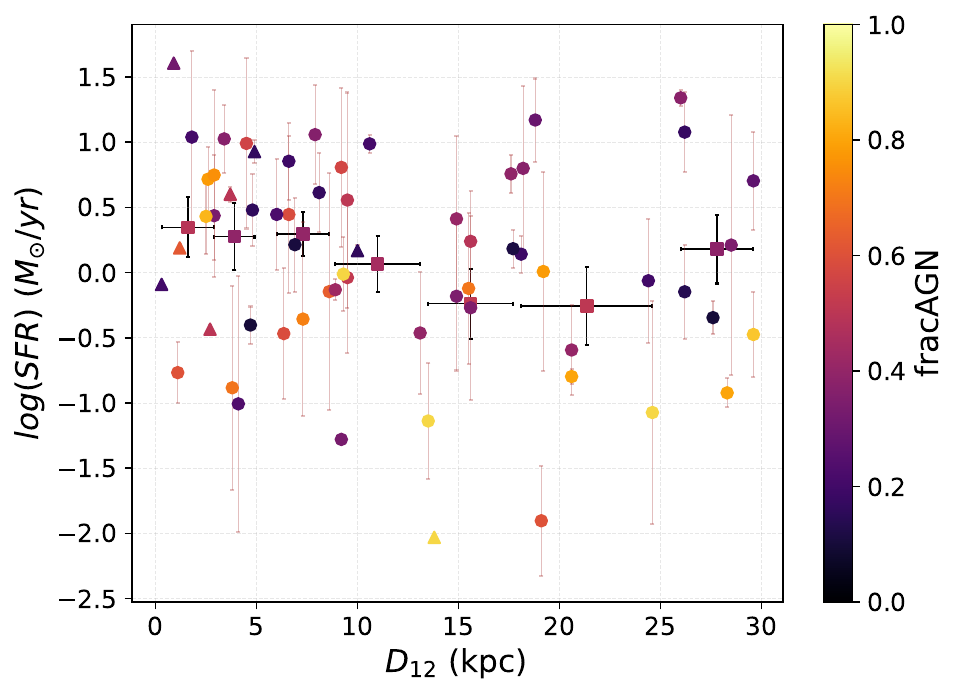}
    \end{subfigure}
    \begin{subfigure}{0.49\textwidth}
        \centering
        \includegraphics[width=\textwidth]{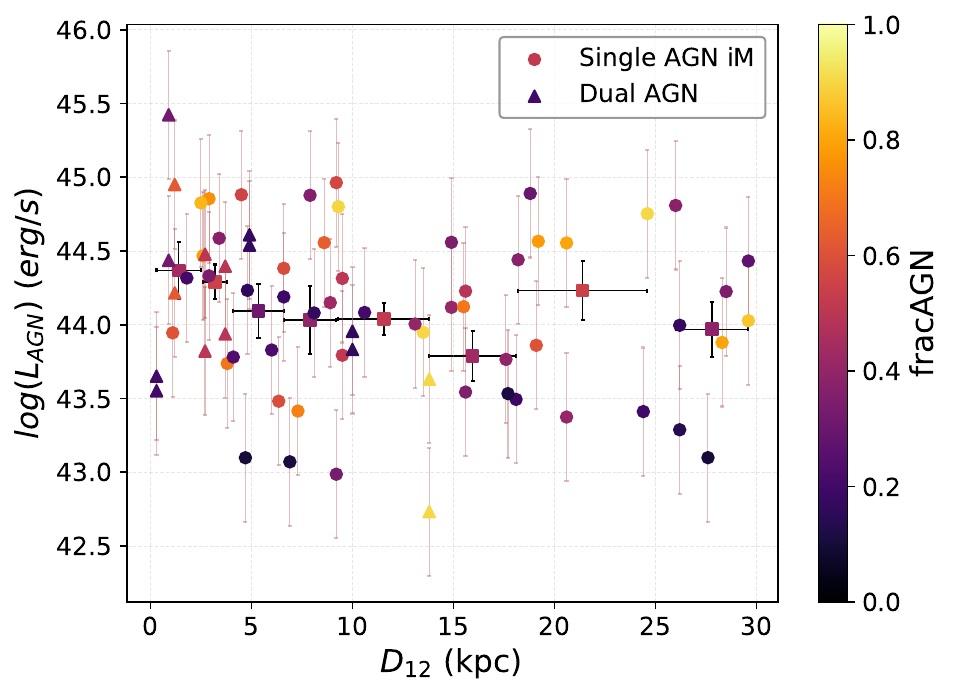}
    \end{subfigure}
    \caption{{\it Left panel}: SFR as a function of the projected nuclear separation. {\it Right panel}: Observed total AGN luminosity as a function of nuclear separation. Both figures are colored by the fraction of AGN luminosity in the IR band. Squares represent the mean value of SFR/$L_{AGN}$ in bins of nine sources.}
    \label{fig:separation}
\end{figure*}

The main motivation of this study is to investigate whether the physical properties of the sample are linked to the evolution of the merger process. We first use the projected nuclear separation to trace the merger stage, where smaller separations correspond to more advanced merger stages. However, as shown in the first column of Figure \ref{fig:spearman}, no correlations were found between projected separation and the physical parameters of the merging systems. This lack of correlation is further illustrated in Figure \ref{fig:separation}, where the sources do not exhibit a distinct pattern or position in the diagram, at least within the observed scatter.

%Details of the nuclear separation measurements are provided in Section \ref{sec:sample}.

This non-relation is unexpected, as simulations predict that such trends should exist \citep[e.g.,][]{2024MNRAS.528.5864B}. Firstly, the assumption that both galaxies in the merging system lie at the same distance from us may not hold. Since projected separations do not account for redshift differences, a merger occurring along the line of sight could potentially overestimate the merger stage. Secondly, galaxies can pass through each other during the merger process, resulting in an increased separation after their initial encounter \citep{1972ApJ...178..623T, 1988ApJ...331..699B}, which can lead to an underestimation of the merger stage. Hence, the lack of a strong correlation between the physical properties of the sources and their projected separation is not unexpected. Therefore, separation and morphological changes must be considered together to assess the merger stage better, as galaxies undergo significant morphological transformations throughout the merging process.

\begin{figure}[b]
    \centering
    \includegraphics[width=\columnwidth]{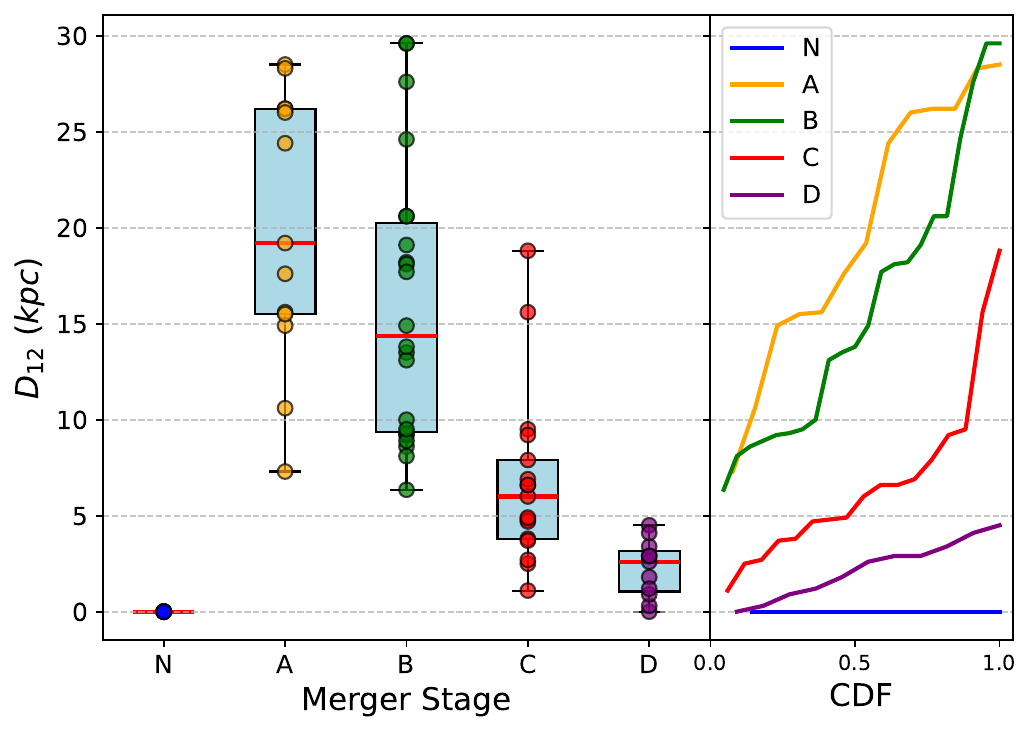}
    \caption{Projected separation as a function of the merger stage (left panel), along with the cumulative distribution function (CDF) for each stage (right panel). Non-merging sources (N) are assigned a projected separation of 0. In the left panel, each rectangle represents the interquartile range (IQR), spanning from the first quartile (Q1) to the third quartile (Q3), while the whiskers extend to the minimum (Q0) and maximum (Q4) values within the range. The red line within each box indicates the median (Q2) of the distribution.} 
    \label{fig: sep}
\end{figure}

%Firstly, the assumption that both galaxies are at exactly the same distance might not hold, as the galaxies could be merging along the radial direction. However, this should have minimal impact since the redshift difference is usually small. Secondly, galaxies could pass through each other during the merger process, resulting in an increased separation after their initial encounter.

Figure \ref{fig: sep} illustrates the relationship between projected separation and merger stage, showing a clear correlation with lower separations concentrated in the more evolved stages. However, separations often overlap, particularly in the A and B stages, which share similar separation ranges. We analyze the properties of our sources in further detail across the merger sequence in Section \ref{sec:dis_merg}.

\section{Discussion} \label{sec:discussion}

In this section, using the results of our CIGALE multiwavelength SED fitting analysis, we study the interplay between AGN and their host galaxies in the context of galaxy mergers (Section \ref{sec:dis_gal}). We then examine how these properties evolve throughout the merging process, using the merger stage as a proxy (Section \ref{sec:dis_merg}). Finally, we discuss potential limitations and caveats of our SED fitting approach (Section \ref{sec:impact}), focusing on the effects of angular resolution differences when measuring fluxes for merging systems, the challenges of modeling dual AGNs within a single SED framework, and the interpretation of fits with unusually high $\chi^2$ values.

\subsection{Co-evolution of the SMBH and host galaxy properties during the merger process} \label{sec:dis_gal}

The connection between SMBHs and their host galaxies is well established \citep{1995ARA&A..33..581K, 1998AJ....115.2285M, 2000ApJ...539L...9F, 2000ApJ...539L..13G, 2009ApJ...698..198G, 2009ApJ...704.1135B}, particularly in the context of galaxy mergers \citep{2006ApJS..163....1H, 2018MNRAS.478.3056B, 2023ApJS..265...37Y}. In this section, we compare our sample of merging galaxies to AGN-hosting systems from the BASS sample and non-AGN systems from the xCOLD GASS sample.  

\subsubsection{Star-Formation Main Sequence} \label{sec: MS}

\begin{figure*}[htbp]
    \centering
    \includegraphics[scale = 0.6]{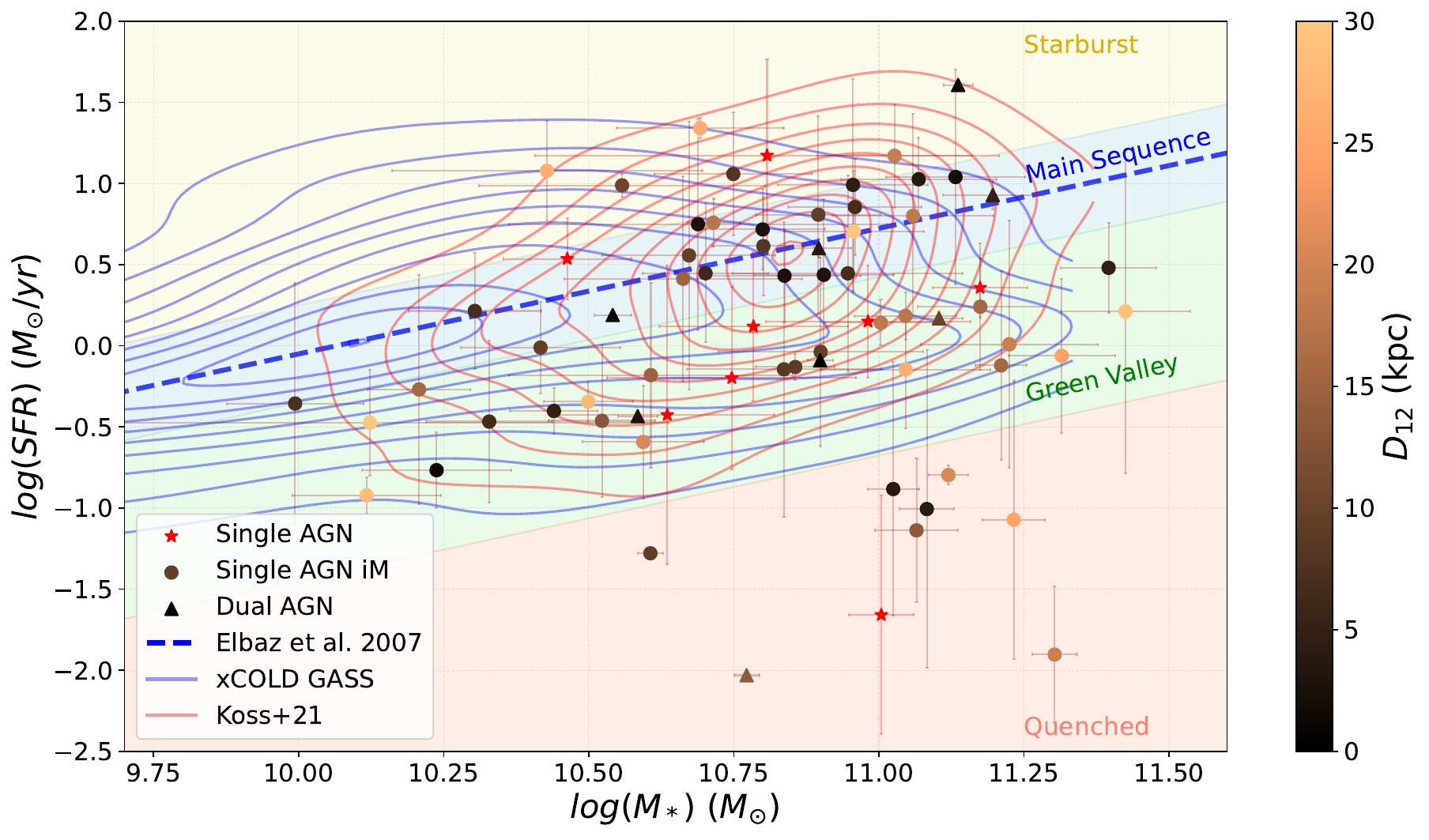}
    \caption{SFR as a function of stellar mass. The blue and red contour maps correspond to the inactive population from xCOLD GASS \citep{2017ApJS..233...22S} and the active population from \cite{2021ApJS..252...29K}, respectively. The dashed blue line marks the local star-forming main sequence relation from \cite{2007A&A...468...33E}, with the background shaded blue region representing a $\pm$0.3 dex width scatter. Additionally, the other background shaded colors highlight different regions of the diagram: salmon for quenched galaxies, green for the green valley, and the upper yellow region for starburst galaxies. The points denote the single AGNs iM, while the triangles represent the confirmed dual AGNs. Both populations are colored by their projected separation. Moreover, red stars represent galaxies where only one nucleus is visible.} 
    \label{fig:mainsequence}
\end{figure*}

Figure \ref{fig:mainsequence} illustrates the position of our sample in the SFR-$M_*$ plane. For comparison, we include a non-AGN population from xCOLD GASS \citep{2017ApJS..233...22S} and the hard X-ray AGN from the BASS survey, for which these values were previously measured \citep{2021ApJS..252...29K}. Our sample does not follow the local main sequence log(SFR) = 0.77$\times$ log($\mathrm{M_*/M_{\odot}}$) - 7.53 from \cite{2007A&A...468...33E}, as seen in Figure \ref{fig:spearman}, and as manifested by a Spearman factor of 0.16. It is important to note that the stellar mass obtained by CIGALE for our sample ranges from $10^{10} M_{\odot}$ to $10^{11.5} M_{\odot}$, following the same distribution as the BASS sample.

%, well aligned with the xCOLD GASS and BASS populations.

%Each point is colored by its projected nuclear separation, with Single AGN iM as dots and dual systems as triangles. Isolated systems with only one nucleus are plotted as red stars. The local main sequence from \cite{2007A&A...468...33E} is also shown

The distribution of our sample in the SFR-$M_*$ plane can be divided into four distinct regions. Starting from the lower SFRs, one group is placed far below the main sequence (9/70), indicating quenched star formation. Next, a separate population appears slightly below the main sequence, tracing the green valley. This region \citep[$\sim$1.0 dex below the MS; e.g.,][]{2014MNRAS.440..889S}, where 31/70 of our sources are located, is consistent with the expectations for galaxies involved in major mergers and also with an active nucleus, as it was found before for AGN in the MANGA survey \citep[e.g.,][]{2021ApJ...923....6J}. The third region lies on the main sequence relation, representing galaxies with normal star formation (22/70). Finally, the top-most group is found above the main sequence (8/70), identified with starburst episodes. By number, these are less likely events, often associated with more extreme ULIRGs in the local Universe (e.g., \citealp{2024ApJ...970...29L}). 

In non-merging galaxies, star formation typically declines gradually over time, with fluctuations around the main sequence before ultimately reaching a quenched state \citep[$\sim$0.3 dex deviation from the MS; e.g.,][]{2016MNRAS.457.2790T}. In contrast, our results suggest that galaxy mergers follow a different evolutionary path: rather than maintaining their position on the main sequence, many merging systems reside in the green valley throughout much of the merging process, indicating that these sources have recently undergone quenching. The distribution of our sources supports this interpretation: most of them are found in this transitional region. As shown by \cite{2014MNRAS.440..889S}, merging galaxies are indeed more likely to lie below the main sequence. However, they are also expected to experience a rapid starburst phase just before coalescence (ULIRGS), resulting in a subset of sources with elevated SFRs \citep{2008ApJS..175..356H,2024ApJ...970...29L}. In \cite{2023ApJS..265...37Y}, using a sample of U/LIRGS from the GOALS survey, the authors reached a similar conclusion, but in the opposite direction. Mergers selected based on their IR emission exhibit elevated SFRs, with positions above the main sequence being the most common scenario. These apparently contradicting results highlight the impact that selection methods can have on the observed properties of the derived samples. %This effect is even more pronounced in our sample, as all of the galaxies host AGNs, which may further influence the suppression of star formation . into the green valley before the quenching occurs

These results suggest that galaxy mergers can significantly alter the position of galaxies from the star-forming main sequence, driving them to different locations in the $\mathrm{SFR}-M_{\odot}$ plane. In particular, for our AGN-merging sample, nuclear activity further complicates this picture, potentially accelerating the depletion of star-forming gas through feedback mechanisms, such as AGN-driven winds and radiative feedback \citep[e.g.,][]{2005Natur.433..604D}, pushing galaxies into the green valley before quenching is complete. To better understand this interplay, we examine the relationship between AGN properties and SFR, focusing on how AGN activity evolves during the merging process.
   
\subsubsection{AGN properties versus SFR} \label{sec: AGN}

\begin{figure*}[htbp]
    \centering
    \begin{subfigure}{0.423\textwidth}
        \centering
        \includegraphics[width=\textwidth]{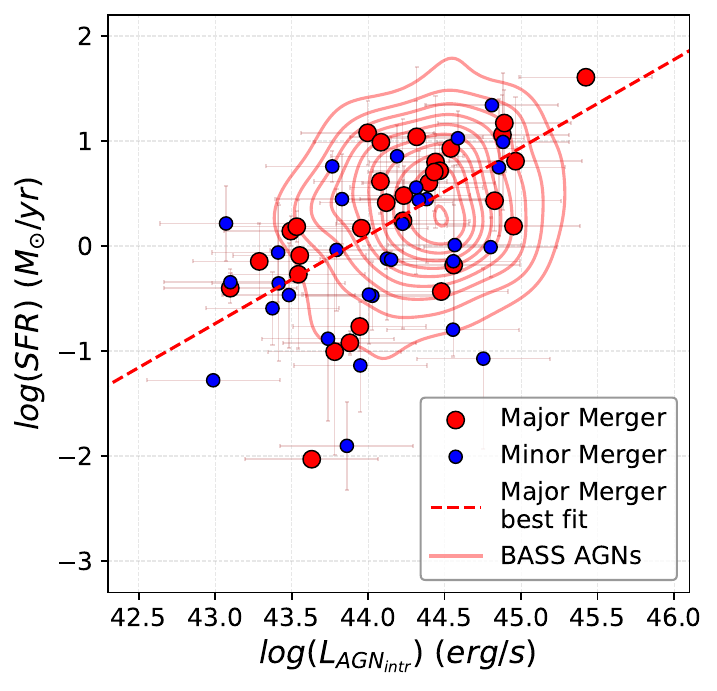}
    \end{subfigure}
    \begin{subfigure}{0.55\textwidth}
        \centering
        \includegraphics[width=\textwidth]{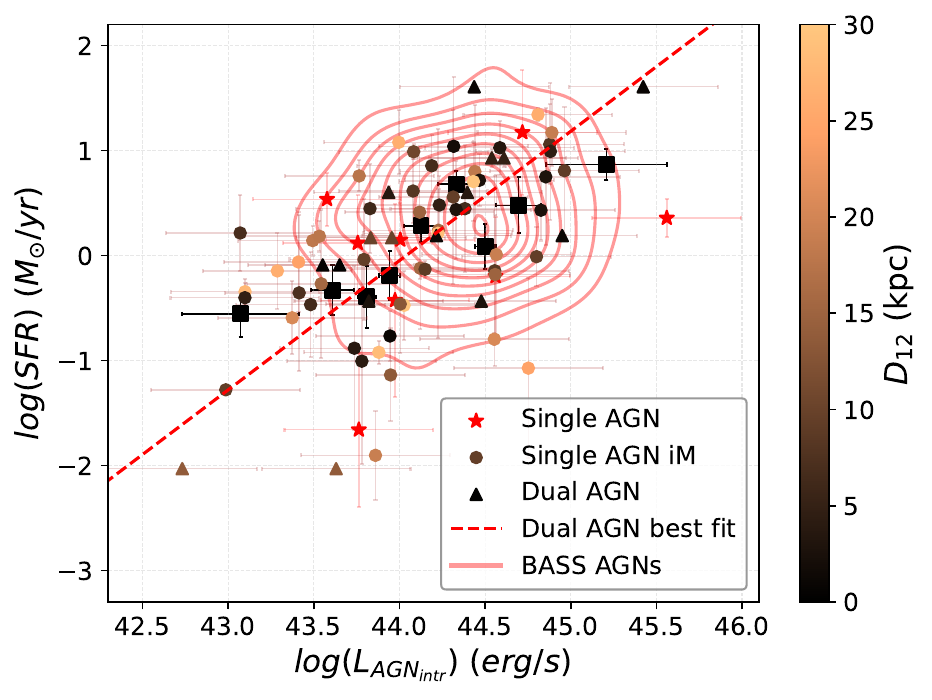}
    \end{subfigure}
    \caption{AGN intrinsic luminosity as a function of the SFR. {\it Left panel}: Red and blue points denote major and minor galaxy mergers, respectively.{\it Right panel}: The red line corresponds to the best linear fit for the dual AGN population. Squares represent the mean value of SFR in bins of nine sources. The symbols and colors of the points are the same as in Figure \ref{fig:mainsequence}. In both panels, the red contours represent the distribution of the sources in the BASS sample.} 
    \label{fig:agnluminosity}
\end{figure*}

The relationship between AGN luminosity and SFR is shown in Figure \ref{fig:agnluminosity}. As indicated in Figure \ref{fig:spearman}, a mild correlation is observed between these two parameters for our merging sample, with a Spearman value of 0.54. The right panel reveals that higher AGN luminosities are associated with higher SFRs in merging systems. In contrast, the non-merging sources from the BASS sample do not exhibit a significant correlation, showing a much lower Spearman coefficient value of 0.18. When separating the mergers by mass ratio, major mergers present a stronger and more significant correlation than minor mergers, as seen in the left panel of Fig.~\ref{fig:agnluminosity}.

Previous studies have highlighted the connection between star formation and AGN activity \citep{2007ApJ...671.1256N, 2009MNRAS.399.1907N, 2010MNRAS.406L..35T, 2014ApJ...780...86E, 2023ApJS..265...37Y}. However, in this hard X-ray selected sample, our merging galaxies and the non-interacting population from \cite{2021ApJS..252...29K} display distinct behaviors. The mild correlation observed in the merging systems suggests that the merger event may simultaneously drive an intense star formation episode in the host galaxy and activate the AGN. This effect is particularly evident for the major mergers in our sample, which show a stronger correlation than the minor mergers sample and the BASS non-merging population.

\begin{figure}[htbp]
    \centering
    \includegraphics[width=\columnwidth]{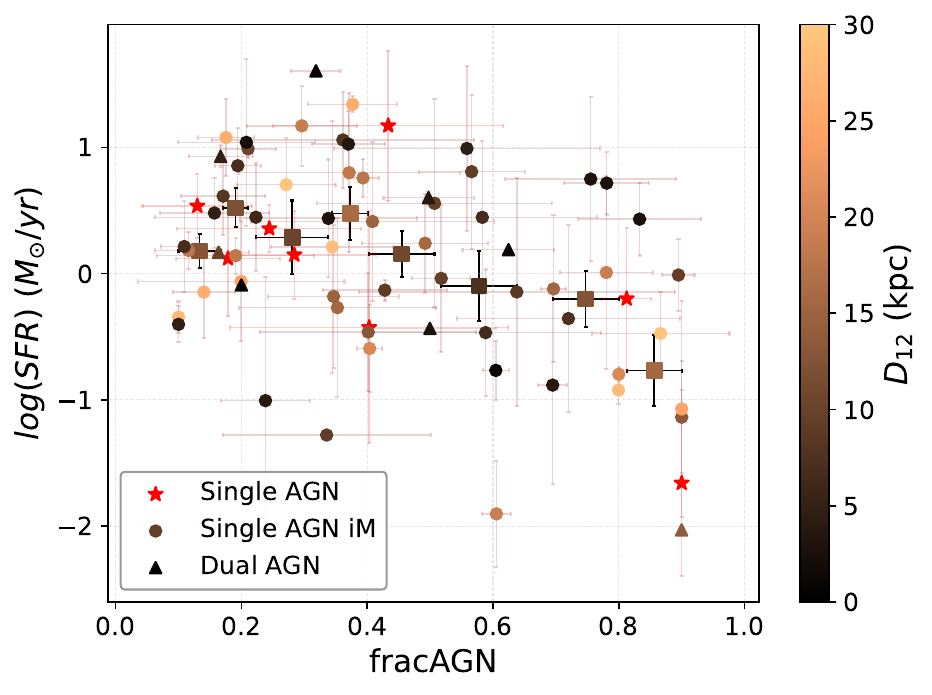}
    \caption{SFR as a function of the fraction of the AGN luminosity in the IR band. Squares represent the mean value of SFR in bins of nine sources. The symbols and colors of the points are the same as in Figure \ref{fig:mainsequence}.} 
    \label{fig:fracagn}
\end{figure}

Nevertheless, when the AGN dominates the IR luminosity of the galaxy, the star formation rate (SFR) slightly decreases, as shown in Figure \ref{fig:fracagn}. Although this trend may partly indicate that AGN activity plays a role in quenching star formation within the host galaxy, the significant scatter in the data suggests that AGN feedback is not a dominant factor. Instead, this trend may be caused by our SED modeling approach, as the IR luminosity in the model is generated by a combination of the AGN and star formation components, which causes them to compete with each other.

%Although this trend could partially result from our SED modeling approach, since the IR luminosity is attributed to a combination of AGN and star formation, the observed correlation between SFR and AGN luminosity suggests that this suppression may not be purely a modeling effect. This correlation, hence, provides evidence that AGN activity may play a role in quenching star formation within the host galaxy.

The impact of AGN feedback, whether negative or positive, is now better constrained, and both mechanisms have been observed in disturbed galaxies \citep{2015ApJ...799...82C}. This confirms that AGN activity and star formation can coexist in the same system. On the one hand, negative AGN feedback becomes significant when the AGN fraction dominates the galaxy's total IR luminosity by expelling or heating the surrounding gas. On the other hand, a higher AGN luminosity does not necessarily imply a lower SFR. In fact, in merging galaxies, we observe the opposite trend. Systems with higher AGN luminosities tend to exhibit enhanced star formation. This correlation suggests that both processes are fueled by the same gas reservoir and driven by the merger process, especially major ones, where material funneled toward the galactic center feeds the SMBH and triggers star formation in the host galaxy \citep{2008ApJS..175..356H, 2024OJAp....7E.121E, 2025OJAp....8E..12E, 2025MNRAS.538L..31F}.

\subsection{Evolution of physical properties during the merger process} \label{sec:dis_merg}

We employ the morphological classification outlined in Section \ref{sec: morphology} to properly study the merger evolution, allowing us to analyze key parameters in different merger stages. By examining the median values within each classification, we uncover trends that are not evident when individual sources are studied independently.

\begin{figure*}[htbp]
    \centering
    \begin{subfigure}{0.49\textwidth}
        \centering
        \includegraphics[width=\textwidth]{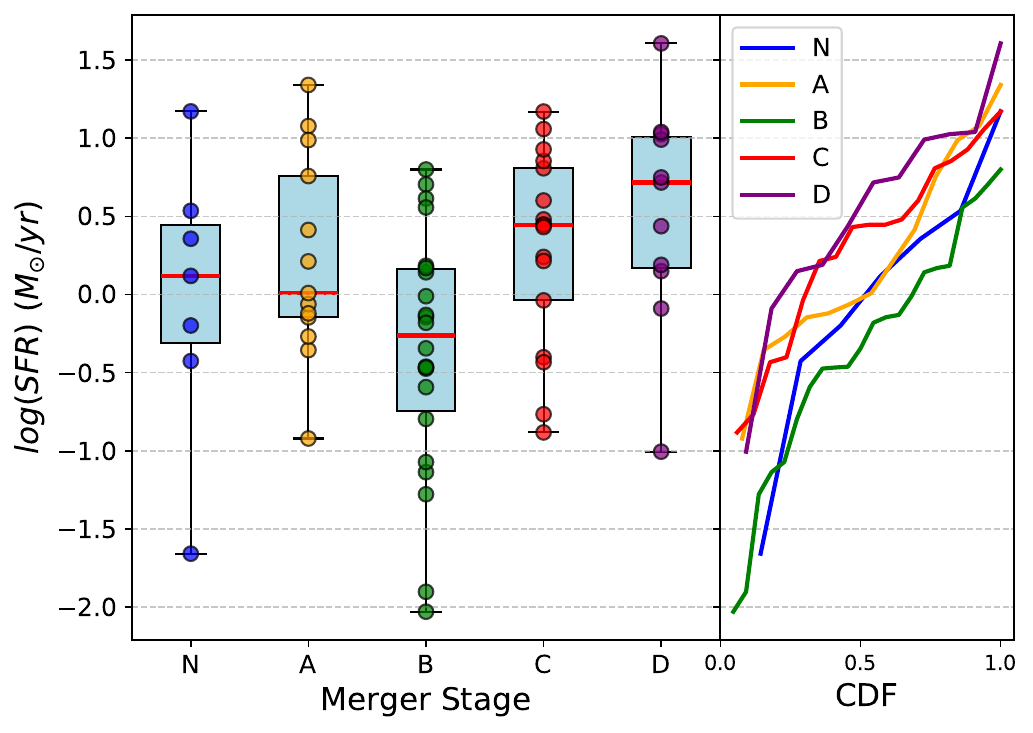}
    \end{subfigure}
    \begin{subfigure}{0.49\textwidth}
        \centering
        \includegraphics[width=\textwidth]{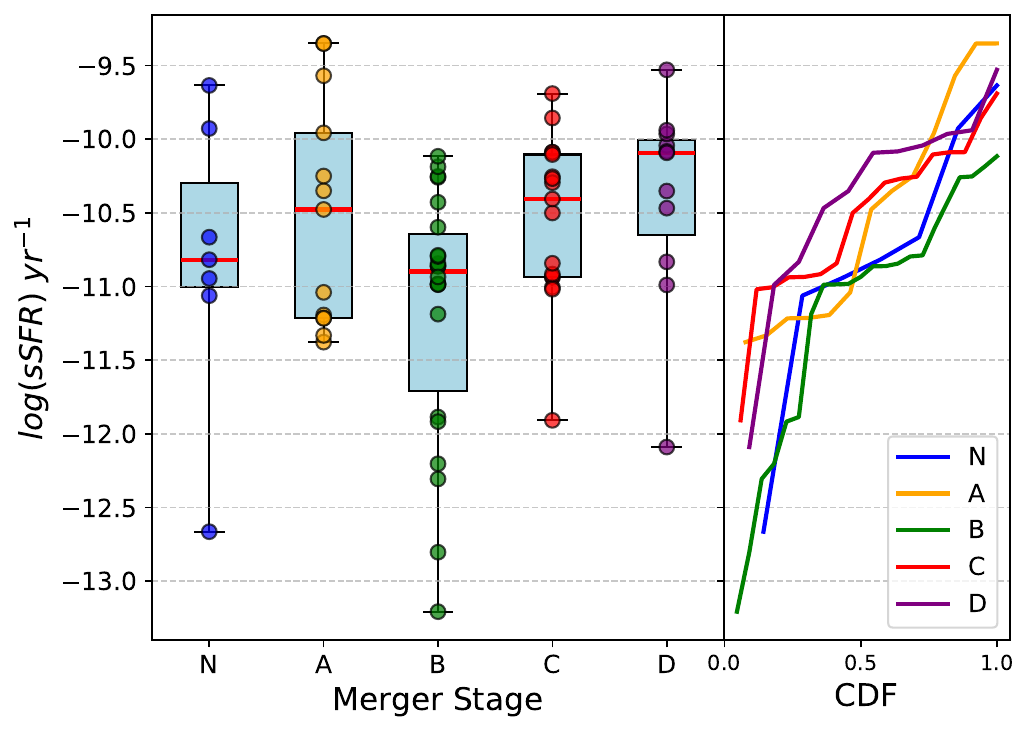}
    \end{subfigure}
    \caption{{\it Left panel}: SFR as a function of the merger stage and  {\it Right panel}: sSFR as a function of the merger stage, along with the cumulative distribution function (CDF) for each stage. Symbols are the same as those defined in Figure \ref{fig: sep}.}
    \label{fig:mergersfr}
\end{figure*}

In terms of the evolution of the SFR, Figure \ref{fig:mergersfr} shows that the highest values occur in the most advanced merger stages, with a factor of 5.1 increase from stage A to stage D. Interestingly, a significant decline (factor of 0.56) is observed between \textit{pre-stage} mergers and \textit{early-stage} mergers. Although this phenomenon could be a genuine physical property of this stage, no other studies have found a similar behavior supporting this result. Moreover, inspection of the plot reveals that the most densely populated values are consistent between stages A and B, with the observed drop largely driven by two outliers with exceptionally low star formation rates.

\begin{figure}[htbp]
    \centering
    \includegraphics[width=\columnwidth]{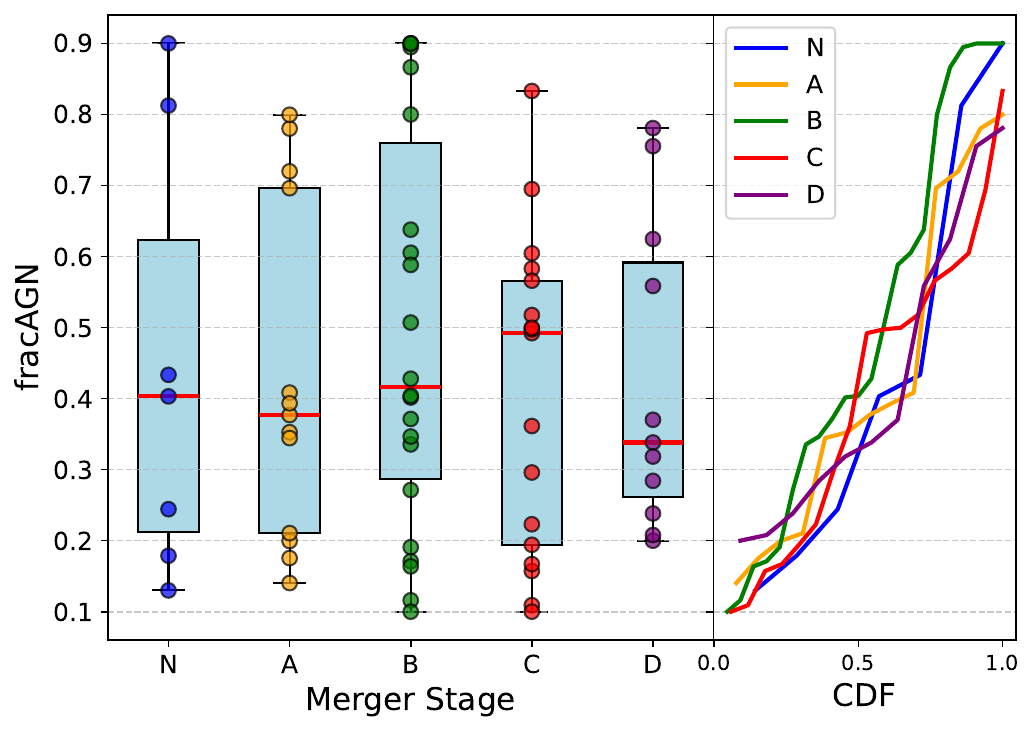}
    \caption{Fraction of the AGN luminosity in the IR band as a function of the merger stage, along with the cumulative distribution function (CDF) for each stage. Symbols are the same as those defined in Figure \ref{fig: sep}.} 
    \label{fig: mergerfrac}
\end{figure}

The dominance of the AGN in the IR luminosity shows no clear correlation with the merger stage, as illustrated in Figure \ref{fig: mergerfrac}. This suggests that the merging process does not significantly impact the ratio between AGN luminosity and SFR, with a wide range of values observed across all stages. In particular, we observe systems ranging from AGN-dominated to SF-dominated at every merger stage, indicating that this ratio remains largely unaffected by the merger progression. Interestingly, in stage D, we see a possible drop in the AGN dominance. This could be consistent with a "blow-out" phase, where powerful AGN feedback clears the nuclear region of gas and dust. However, given the limited statistics, we cannot confirm this interpretation, and we therefore leave it as a possibility.

 Particularly in the D stage, we can observe a possible drop, that could correspond to a blow-out phase, evacuating the nuclear region and pushing the SF in the outskirts. However, we do not have the necessary evidence to state it, so we left it there as a caveat.

\begin{figure}[htbp]
    \centering
    \includegraphics[width=\columnwidth]{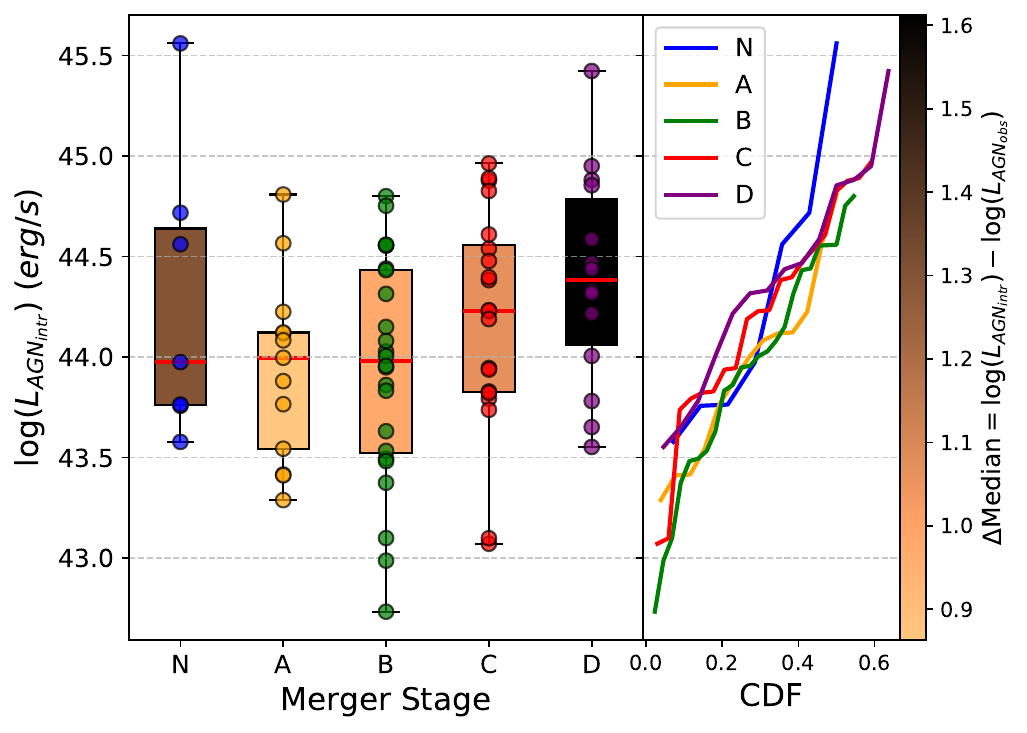}
    \caption{AGN intrinsic disk luminosity as a function of the merger stage. Each box is colored by the difference between the median AGN intrinsic disk luminosity and the median AGN observed disk luminosity. Symbols are the same as those defined in Figure \ref{fig: sep}.} 
    \label{fig:mergagnL}
\end{figure}

However, the luminosity due to the AGN accretion disk presents a trend with the merger stage, as shown in Figure \ref{fig:mergagnL}. On average, the luminosity progressively increases with more advanced merger stages, exhibiting a 2.4-fold increase from stage A to stage D, which indicates that the merger process accelerates the nuclear accretion rate. This result is consistent with theoretical expectations and has been previously reported in various observational studies \citep{2012ApJ...746L..22K, 2023ApJS..265...37Y} and simulations \citep{2015MNRAS.447.2123C, 2023MNRAS.524.4482B}. Additionally, the plot color-coded highlights the difference between the intrinsic and observed disk AGN luminosity, revealing that more evolved merger stages are associated with higher extinction levels by a factor of 2.8, as previously reported by \cite{2017MNRAS.468.1273R, 2021MNRAS.506.5935R}.

As shown in the figures described above, the AGN luminosity and the SFR increase in the final stages of the mergers. The most significant difference between the early stages (A and B) and the late stages (C and D) is the degree of disturbance in their disks \citep{2013ApJS..206....1S}, which likely triggers both processes. This suggests that the same mechanism fuels star formation and black hole accretion as the merger progresses, driving the system toward its final shape and structure. 

\subsection{SED Fitting Results}

In this section, we discuss the implications of the SED fitting methodology for our analysis, particularly the impact of using merging galaxies in a SED fitting code (section \ref{sec:impact}), and a more detailed study of the poorer fits in our sample (section \ref{sec:badfits}).

\subsubsection{Impact of angular resolution differences in the merger's SED} \label{sec:impact}

As discussed in Section \ref{sec: fluxes}, resolving both galaxies individually for SED fitting is not feasible in our study due to the limitation of the spatial resolution of bands such as WISE W4 and PACS. Consequently, we derive the physical parameters for the two galaxies as a single system. This approach has been employed in previous studies. In particular, \citet{2023ApJS..265...37Y} quantified the impact of merging two galaxies into a single SED fit for non-active systems and single AGNs in mergers. Their findings indicate that while the host galaxy properties remain unchanged, the AGN luminosities differ by approximately 0.2 dex, a small effect that is unlikely to impact our results significantly. %Unfortunately is not possible to measure that in our work.

Roughly, this method produces a luminosity-weighted average of both galaxies; however, its effects on the physical parameters derived from the SED fitting may depend on the merger stage. In the less evolved mergers (A and B stages), where both disks remain independent, the individual properties of each galaxy may still influence the results. However, their physical properties would be mixed in later stages (C and D), where the disks have already merged. 

\begin{figure}[htbp]
    \centering
    \includegraphics[width=\columnwidth]{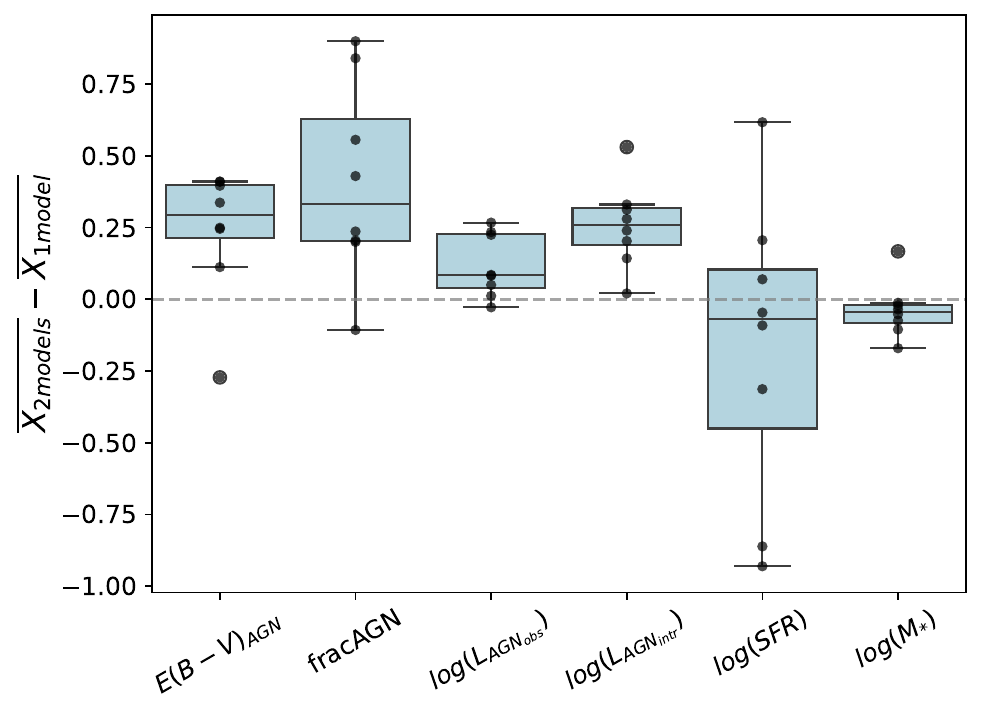}
    \caption{Comparison of physical properties of dual AGNs measured using one versus two AGN components. Each point represents the difference between the values obtained with two AGN components and those derived from a single-component model. For AGN luminosities in the two-component model, the total value corresponds to the sum of both luminosities.} 
    \label{fig:dual_test}
\end{figure}

From the AGN perspective, single AGNs in mergers are unlikely to be affected, given that we model only a single AGN. However, for confirmed dual AGNs, applying a single-component AGN fit could introduce inaccuracies due to their potentially distinct characteristics, such as varying levels of obscuration, accretion rates, and torus geometries. As presented in Section \ref{sec: sed}, we address this issue by incorporating two AGN components in CIGALE for a more realistic modeling of the combined SED. Figure \ref{fig:dual_test} illustrates the impact of using a single-component AGN model instead of two separate AGN components. In general, SFR tends to be overestimated, whereas the AGN properties are underestimated in the single-component model. This is due to degeneracies in IR emission, where SFR and AGN luminosity both contribute in a similar wavelength range. Nevertheless, our multiwavelength approach must address this issue, particularly through X-ray observations, which may help determine the intrinsic AGN luminosity for most sources. Given this, the overall differences remain minimal. 

Including longer wavelength fluxes would further improve the SED fitting by extending the coverage. Nevertheless, \cite{2022ApJ...938...87K} found a strong correlation between X-ray and millimeter emission in the BASS sample, suggesting that additional data in that range may not be necessary for our study, given that the information is already there. Moreover, radio data might be complicated due to the radio loud/quiet dichotomy \citep{2008MNRAS.387..856Z, 2012ApJ...759...30B}, and hence such a study, incorporating radio data, is beyond the scope of this work.

In the near future, repeating this analysis at higher spatial resolutions across the entire wavelength range, particularly in the IR using telescopes such as JWST, would be the most effective way to separate the emission from the participating galaxies. However, currently, no instruments cover the range between 30 $\mu m$ and 400 $\mu m$ with the required spatial resolution to separate galaxies at projected distances below $\sim$12" (WISE4 resolution), and ideally resolve nuclear regions, as achieved in the UV, optical and NIR with HST, JWST, AO ground-based imaging, or evenly in X-rays with Chandra. This limitation implies that we must wait for future observational capabilities to fully resolve these systems spatially in the entire wavelength spectrum.

\subsubsection{Analysis of Fits with Large Residuals} \label{sec:badfits}

In Section \ref{sec: results}, we identified four sources with exceptionally large $\chi^2$ values that deviate significantly from the expected distribution. Their SEDs are presented in the Figure \ref{fig: sed_examples}. These poor fits likely arise from systematic issues. 
For BAT ID 13 and BAT ID 489, we suspect that the high $\chi^2$ values result from template incompleteness, as neither source exhibits unusual behavior, and no specific filter dominates the residuals. BAT ID 72 has the highest $\chi^2$, and is classified as a dual AGN. While its fluxes follow a consistent shape, the model fails to reach the required luminosity, even when incorporating two AGN components. Since the X-ray bands primarily constrain AGN luminosity, the observed IR flux deficit may indicate that the relatively bad fit can be due to an incomplete template for the star formation parameters. Finally, BASS ID 329 has most of its $\chi^2$ arising in the NUV filter, suggesting that the poor fit may be due to inaccurate photometry in that band. 

%These examples highlight the limitations of current templates and data quality in a few complex cases.

\section{Conclusions} \label{sec: conclusion}

Using a sample of 72 AGN-hosting galaxy mergers from the BASS survey, we conducted SED fitting with self-measured fluxes with homogeneous apertures across all the included filters, spanning from X-rays to FIR. This multiwavelength approach ensures consistency across the entire spectral range, allowing us to accurately constrain the physical properties of both the host galaxies and their central SMBHs throughout the merger process. Our study provides new insights into the interplay between AGN activity, star formation, and the evolution of merging systems in the context of local universe ($z<$0.1) AGNs. The main conclusions of our analysis are as follows:

\begin{enumerate}[topsep=5pt, itemsep=0pt, leftmargin=12pt]
    \item Merging galaxies with active nuclei are not predominantly found around the star-forming main sequence. Instead, they span all possible cases: quenched systems, green valley, main sequence, and starburst galaxies. Most systems are found below the main sequence, in the green valley region.
    \item The merger process can enhance AGN activity and star formation, finding a correlation between these two physical parameters for our interacting systems. We also find an opposite scenario for non-merging AGNs, which do not present a clear connection. This trend is expected since these sources are mainly fueled by stochastic processes.
    \item The strength of the AGN–star formation relation depends on the merger mass ratio, where major mergers show a clear correlation between AGN luminosity and SFR, whereas minor mergers do not. This highlights the critical role of major galaxy mergers in setting up the connection between SMBH growth and galaxy evolution.
    \item The dominance of the AGN at IR wavelengths could suggest a quenching of star formation. However, this trend may be driven by our SED fitting approach, where the IR luminosity is distributed between the AGN and star formation components, leading to a competition between them.
    \item The projected nuclear separation is not necessarily a reliable tracer of the merger stage. Instead, other tracers, such as visual studies of morphological features, can help to assess the coalescence stage. %is important to use other observational features to assess the stage of the coalescence along with the separation.  Nuclear separation is easily readily available which can be improved once additional informations such as mophology es incorporated.
    \item Incorporating this morphological analysis, we find that late-stage mergers present higher SFR, AGN luminosity, and obscuration, supporting the idea that the final stages of mergers enhance the SMBH growth along with the host galaxy evolution.
    \item Fitting two AGNs with only one component generates discrepancies in the accretion rate and SFR in the galaxy. Using two components obtains more consistent values, making the results more reliable.
\end{enumerate}

Our work indicates that galaxies undergoing a merger process experience significant changes in their physical properties, which continue to evolve as the merger progresses. Future studies at relatively long wavelengths will require high angular resolution observations to gain deeper insights into the role of mergers in shaping the co-evolution of SMBHs and their host galaxies. Separately resolving both merging components will be essential to understanding the individual contributions of each galaxy to the overall system, particularly in the final stages of the coalescence.

\begin{acknowledgments}
We thank the anonymous referee for a very positive and constructive review that, in our opinion, significantly improved the quality and presentation of this article. MT, ET, AR, MB, FEB, RJA, and CR acknowledge support from the ANID CATA-BASAL program FB210003. ET and FEB acknowledge support from FONDECYT Regular - 1190818, 1200495, 1241005, and  1250821. FEB acknowledges support from the ANID Millennium Science Initiative Program ICN12\_009. ET would like to thank the hospitality of the North American ALMA Science Center (NAASC) at NRAO during his sabbatical stay in 2022, where a significant fraction of this work was carried out. M.K. acknowledges support from NASA through ADAP award 80NSSC22K1126. This work was initiated in part at the Aspen Center for Physics, which is supported by National Science Foundation grant PHY-2210452. IMC acknowledges support from the FONDECYT Postdoctorado program 3230653. CR acknowledges support from Fondecyt Regular grant 1230345 and the China-Chile joint research fund. MB acknowledges support from the French government through the France 2030 investment plan managed by the National Research Agency (ANR), as part of the Initiative of Excellence of Université Côte d’Azur under reference number ANR-15-IDEX-01. RJA was supported by FONDECYT grant number 1231718. KO acknowledges support from the Korea Astronomy and Space Science Institute under the R\&D program (Project No. 2025-1-831-01), supervised by the Korea AeroSpace Administration, and the National Research Foundation of Korea (NRF) grant funded by the Korea government (MSIT) (RS-2025-00553982). MS acknowledges financial support from the Italian Ministry for University and Research, through the grant PNRR-M4C2-I1.1-PRIN 2022-PE9-SEAWIND: Super-Eddington Accretion: Wind, INflow and Disk-F53D23001250006-NextGenerationEU.

%support from the ANID BASAL project FB210003. This work was

\end{acknowledgments}

\clearpage

\appendix
\restartappendixnumbering

\section{Mock Analysis} \label{app: mock}

\begin{figure*}[htbp]
    \centering
    \resizebox{0.81\textwidth}{!}{ % Adjust width scale, height will scale proportionally
    \begin{minipage}{\textwidth}
        % First row
        \begin{subfigure}{0.45\textwidth}
            \centering
            \includegraphics[width=\textwidth]{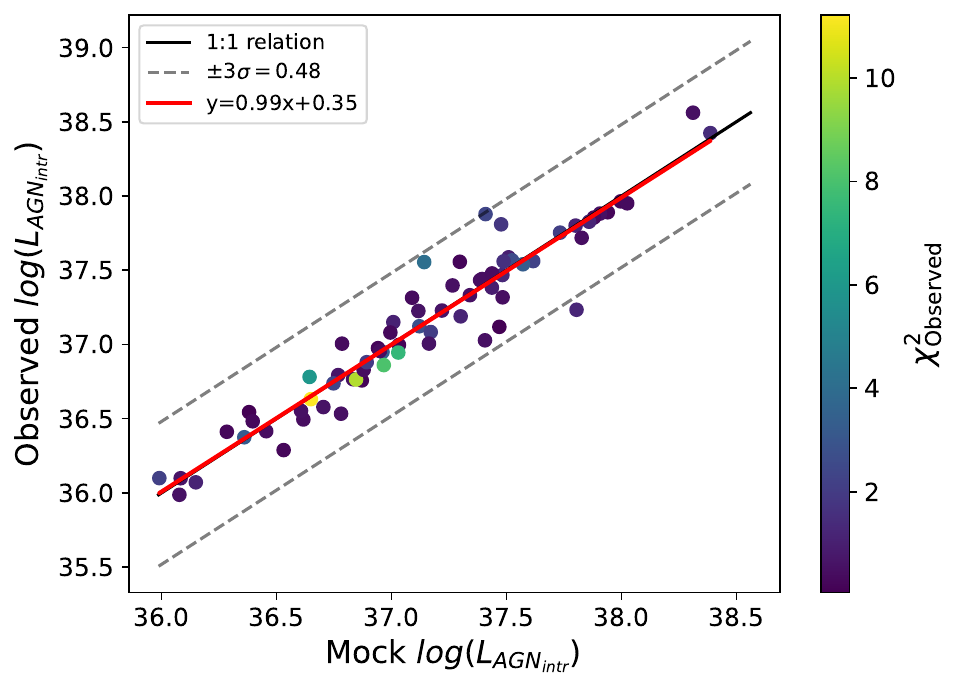}
        \end{subfigure}
        \begin{subfigure}{0.45\textwidth}
            \centering
            \includegraphics[width=\textwidth]{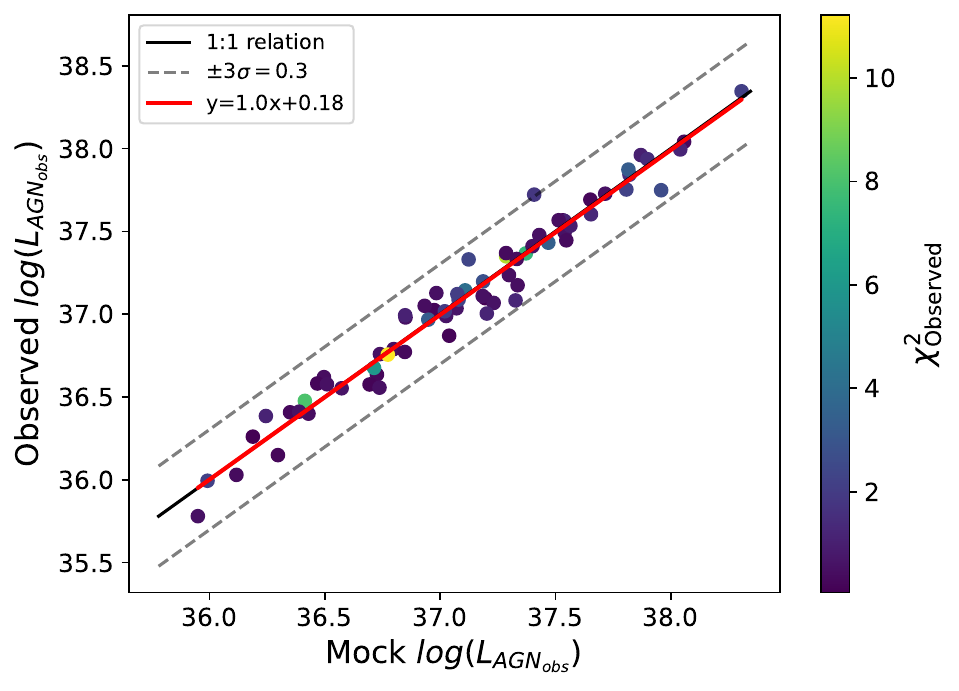}
        \end{subfigure}
        
        % Second row
        \begin{subfigure}{0.45\textwidth}
            \centering
            \includegraphics[width=\textwidth]{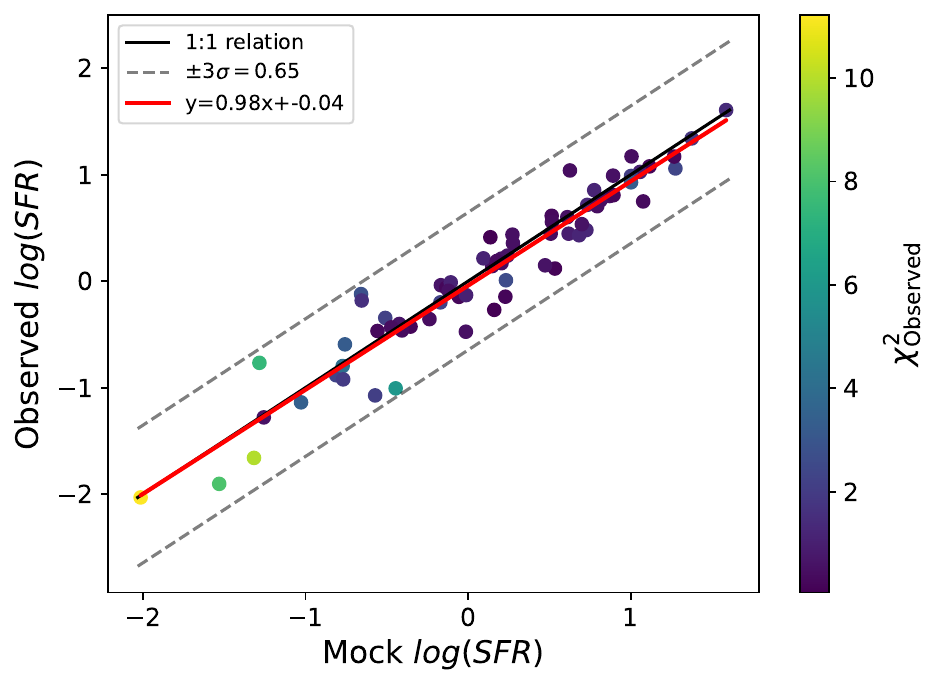}
        \end{subfigure}
        \begin{subfigure}{0.45\textwidth}
            \centering
            \includegraphics[width=\textwidth]{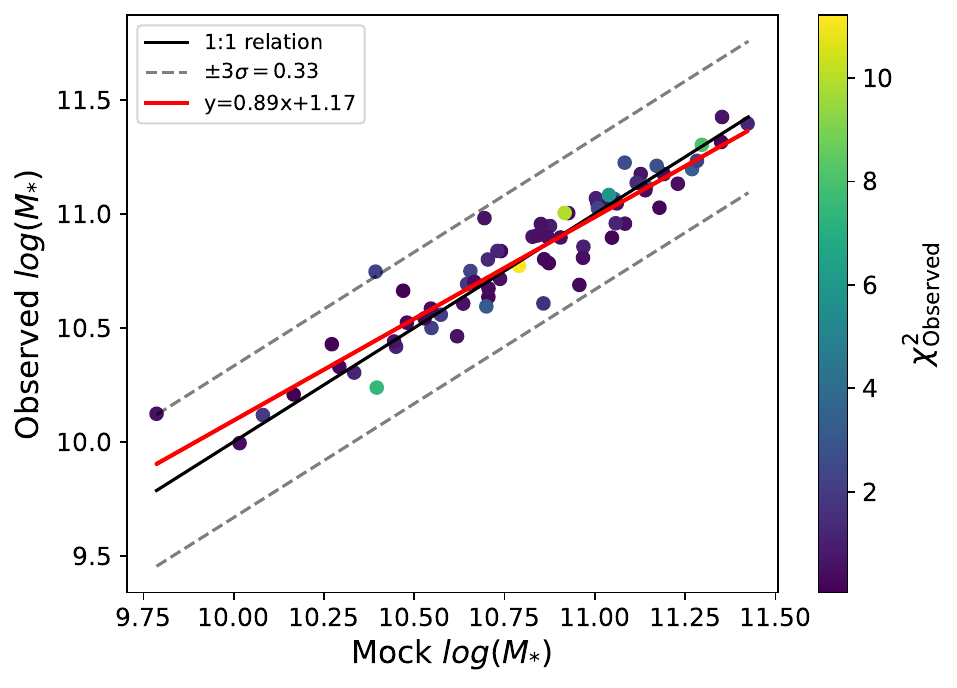}
        \end{subfigure}
    
        % Third row
        \begin{subfigure}{0.45\textwidth}
            \centering
            \includegraphics[width=\textwidth]{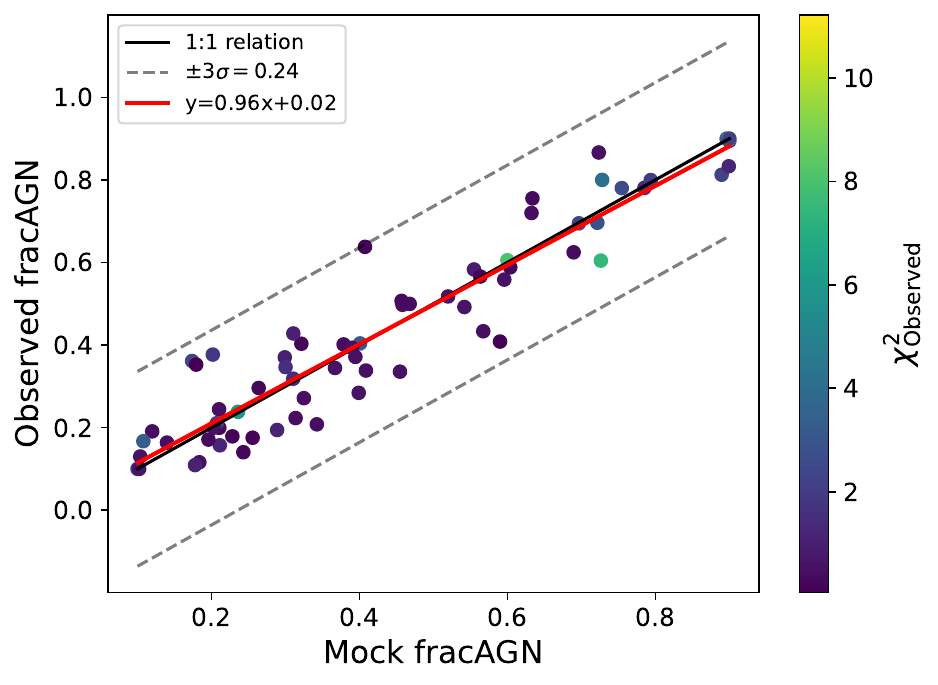}
        \end{subfigure}
        \begin{subfigure}{0.45\textwidth}
            \centering
            \includegraphics[width=\textwidth]{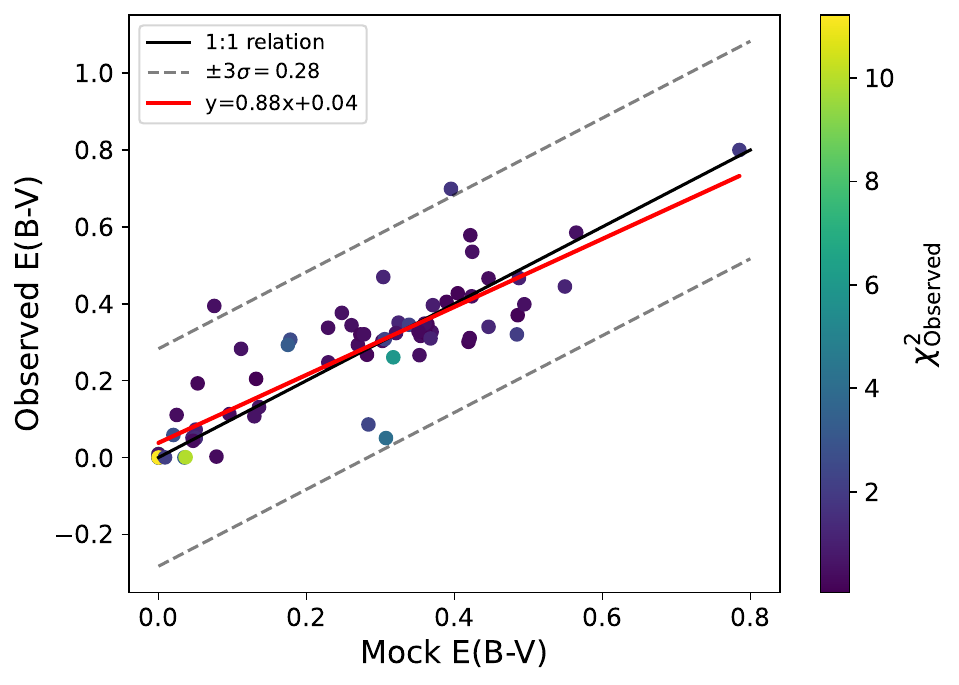}
        \end{subfigure}
    \end{minipage}
    }
    
    \caption{Mock analysis for the key physical parameters used in this study. The x-axis shows the values recovered with the mock fluxes, while the y-axis displays the values obtained from the observed SEDs. Each point is colored by the reduced $\chi^2$ value of the observed fit. The black diagonal line indicates the 1:1 relation, the red line shows the best linear fit, while the dashed lines show the $3\sigma$ range.}
    \label{fig:mock}
\end{figure*}

To test the robustness of our results, we performed a mock analysis to evaluate potential degeneracies in the derived physical properties. Figure \ref{fig:mock} presents the outcome of this analysis, including the 1:1 relation, the 3 $\sigma$ range, and the best-fit linear regression. The AGN luminosities show a strong and tight correlation, with a 1:1 line and the best-fit line overlapping, and no outliers outside the 3$\sigma$ interval.  Regarding the SFR, it also exhibits low scatter and closely follows the 1:1 relation. In contrast, the $M_*$ shows two outliers beyond the 3$\sigma$ range, which affects the quality of the fit, though the overall scatter remains small. The AGN fraction follows the expected trend but with a higher scatter. Finally, the AGN obscuration presents the highest dispersion and a less significant fit, suggesting possible degeneracies in these measurements. Nevertheless, most points still lie close to the 1:1 relation, supporting the overall reliability of the fitting process.

\restartappendixnumbering

\section{Dual AGN SED}

\begin{figure*}[htbp]
    \centering
    \begin{subfigure}{0.49\textwidth}
        \centering
        \includegraphics[width=\textwidth]{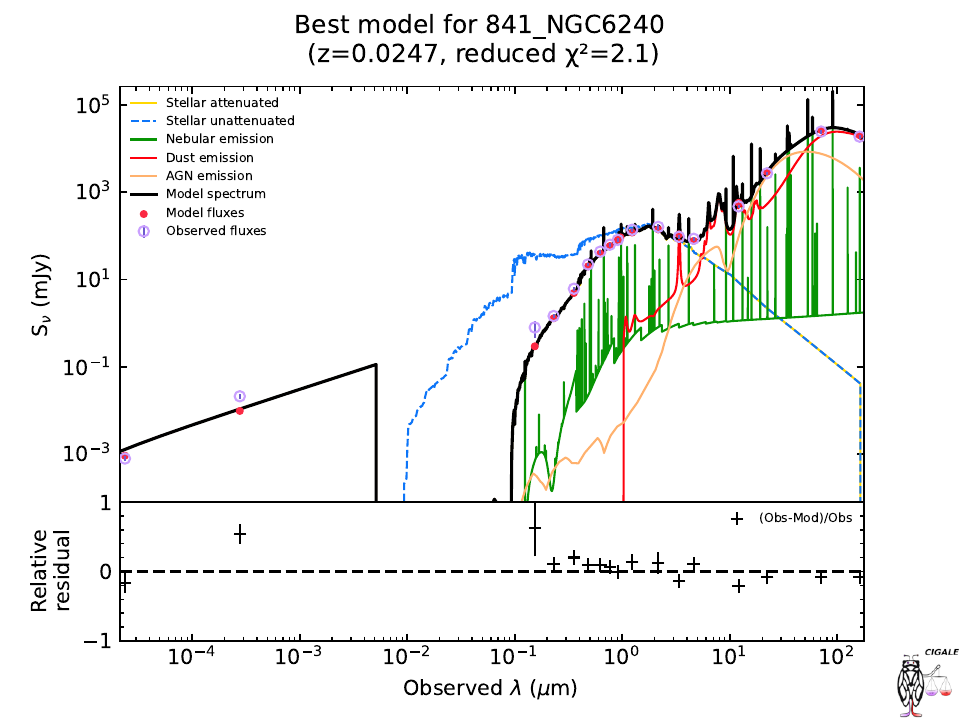}
    \end{subfigure}
    \begin{subfigure}{0.49\textwidth}
        \centering
        \includegraphics[width=\textwidth]{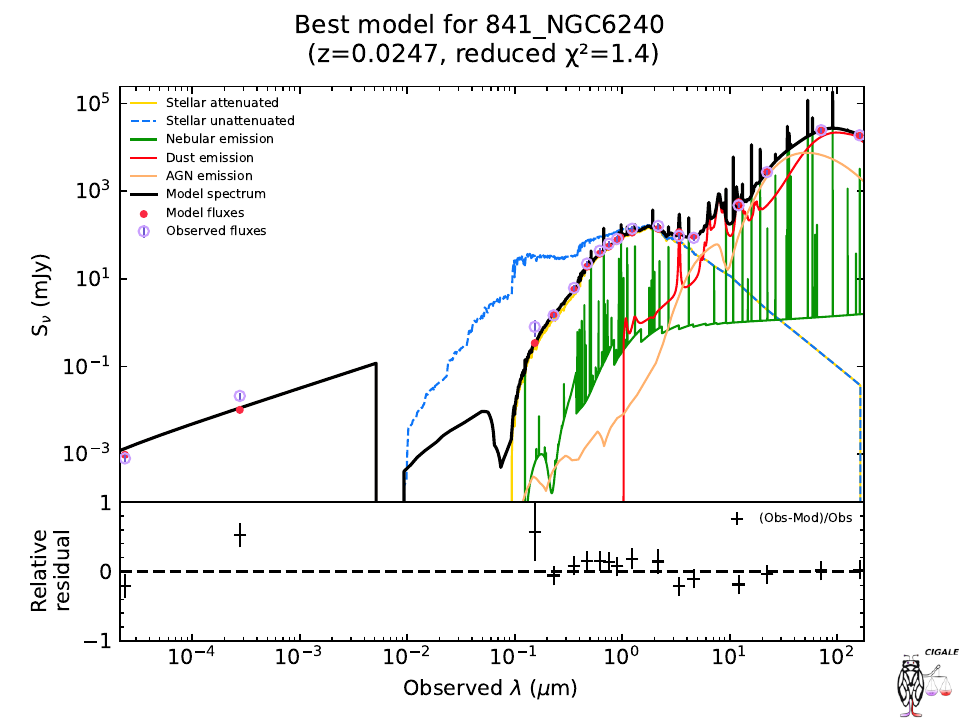}
    \end{subfigure}
    \caption{Best-fit SED for BAT ID 841 using a 1-AGN model (left) and a 2-AGN models (right). The 2-AGN model provide better fit while preserving the same host-galaxy parameters.}
    \label{fig: sed_duals}
\end{figure*}

We applied the 2-AGN model approach to the eight confirmed dual AGN (excluding the 9th dual AGN, BAT ID 497, which did not converge to a reliable solution). On average, the fits improve by 49\% with standard deviation of 14\%. As an illustrative example, Figure \ref{fig: sed_duals} shows the case of BAT ID 841, where the 2-AGN fit achieves lower reduced $\chi^2$ compared to the 1-AGN model. while keeping the same host-galaxy parameters fixed.

\restartappendixnumbering

\clearpage

\bibliography{references_P1}{}
\bibliographystyle{aasjournal}

\end{document}